\documentclass[preprint2]{aastex}
\shorttitle{SSA13 Radio/Optical Catalog}
\shortauthors{Fomalont et al.}
\begin{document}

\title{The Radio/Optical Catalog of the SSA13 Field}

\author{E.~B.~Fomalont, K.~I.~Kellermann}
\affil{National Radio Astronomy Observatory, Charlottesville, VA 22903}
\email{efomalon@nrao.edu, kkellerm@nrao.edu}

\author{L.~L.~Cowie$^1$, P. Capak$^{1,2}$, A.~J.~Barger$^{1,3,4}$}
\affil{$^1$Institute for Astronomy, Honolulu, HI 96822 \\
$^2$California Institute of Technology, Pasadena, CA 91125 \\
$^3$University of Wisconsin-Madison, Madison, WI 53706 \\
$^4$University of Hawaii, Honolulu, HI 96822}
\email{cowie@ifa.hawaii.edu, capak@astro.caltech.edu, barger@astro.wisc.edu}

\author{R.~B.~Partridge$^1$, R.~A.~Windhorst$^2$, E.~A.~Richards$^3$}
\affil{$^1$Haverford College, Haverford, PA 19041 \\
$^2$Arizona State University, Tempe, AZ 85287 \\
$^3$Talledega College, Talladega, AL 35160 }
\email{bpartrid@haverford.edu, Rogier.Windhorst@asu.edu, erichards@talladega.edu}

\begin{abstract}
We present a 1.4-GHz catalog of 810 radio sources (560 sources in the
complete sample) found in the SSA13 field (RA=$13^h12^m$,
DEC=$42^\circ 38'$).  The 1.4-GHz radio image was obtained from a
91-hour VLA integration with an rms noise level of 4.82 $\mu$Jy/beam
at the field center.  Optical images in r-band (6300A) and z-band
(9200A) with three-$\sigma$ detection magnitudes of 26.1 and 24.9,
respectively, were obtained from three observing nights on the 8-m
Subaru Telescope.  We find that $88\pm 2\%$ of the radio sources are
identified with an optical counterpart.  There is significantly more
reddening for the optical counterparts that are fainter than 24-mag,
probably caused by the somewhat larger redshift of these faint
galaxies.  The radio and optical parameters are tabulated, and source
morphologies are displayed by radio contours overlaying optical
false-color images.  The brightness distributions show a wealth of
complexity and these are classified into a small number of categories.
About one-third of the radio sources are larger than $1.2''$ and their
orientation is often similar to that of the associated galaxy or
binary-galaxy system.  Radio emission is sometimes located outside of
the nuclear regions of the galaxy.  The density of sources in the
SSA13 field above $75~\mu$Jy is 0.40 (arcmin)$^{-2}$, with a slope of
$-2.43$ in the differential counts.  This source density is about a
factor of two higher than that in the HDF-north.  The radio spectral
index may steepen for sources below $75~\mu$Jy and is consistent with
the difference in the slope of the source counts observed between 1.4
GHz and 8.4 GHz.  We estimate that at most 40\% of the microJansky
radio sources are dominated by AGN processes while the remainder are
mostly the consequence of star formation and associated supernovae
activity.

\end{abstract}
\keywords{Galaxies:Active; Galaxies:Starburst; Radio
Continuum:Galaxies}

\section{Introduction}

    One of the fundamental problems in astrophysics is the formation
and evolution of stars, galaxies and black holes, and their properties
as a function of cosmic time.  Many radio sources stronger than about
1 mJy are associated with luminous galaxies with an active galactic
nucleus (AGN) that contains a super massive black hole (SMBH).  The
energy generated by interactions near the SMBH produces most of the
observed electromagnetic emission, some of which extends up to several
megaparsecs from the AGN.  These extended active galaxies are often
referred to as `classical double radio sources'.  With the increased
sensitivity of radio observations, now extending to microJansky levels, a
different population is detected \citep{mit85,oor85}.  These sources
are dominated by emission from massive star formation in a galaxy or
in groups of galaxies, have redshifts typically between 0.5 and 2.0,
and appear to evolve strongly with cosmological time.  AGN activity,
although not as dominant as in the brighter radio sources, may still
be an important ingredient in the formation and evolution of the
microJansky sources.

    Since these processes have a complicated signature throughout the
electromagnetic spectrum, multi-band observations are needed to
understand the physical phenomena and evolution.  Sensitive radio
observations are a crucial part of the data base for many reasons:
surveys provide an unbiased sample of a large number of objects
in a relatively small field of view; the radio position accuracy of
$\approx 0.2''$ is crucial in finding sub-mm/optical/X-ray
counterparts; the radio emission is not significantly absorbed by dust;
the FIR/radio correlation is an indication of redshift; and finally, the
radio source spectrum, angular size and the precise location with
respect to the galaxy nucleus can often distinguish between AGN
processes and starforming processes.

    The Small Selected Area 13 (SSA13) field is one of the Hawaii Deep
Fields that has been studied at a variety of wavelengths, comparable
to studies of the Hubble Deep Field (HDF). The first deep VLA radio
observations of SSA 13 were made at 8.4 GHz \citep{win95,fom02} where
about 50 sources were detected in the relatively small part of the
SSA13 region that was imaged.  In this paper we report on additional
deep VLA observations at 1.4 GHz over the entire SSA13 field to a
detection level of $25.8~\mu$Jy/beam.  Our deep r-band and z-band images
from the 8-m Subaru Telescope allows us to present a radio/optical catalog of
over 800 radio sources and describe their radio and optical
properties.

    The radio observations, data reduction and images are described in
\S 2, and the optical observations in \S 3.  The radio/optical catalog
is given in \S 4 in tabular and graphical form, including descriptive
notes for the more interesting objects.  The identification rate, the
radio/optical morphologies, and the radio properties are discussed in
\S 5, \S 6 and \S7, respectively.  A summary is given in \S 8.
Further discussion of the properties of the sources in the SSA13 field
is given elsewhere \citep{cow04}.

\section{The Radio Data}

\subsection{The Observations and Reductions}
    The SSA13 observations were made with the VLA at 1.4 GHz for a
total of 104 hours on nine separate observing days, as shown in Table
1.  Excluding calibration observations, 91 hours of data were obtained
on the SSA13 field.  For eight of the days, the VLA was in the
A-configuration (a maximum baseline of 35 km), and on one day we
observed in the B-configuration (maximum baseline of 10 km) in order
to increase the image sensitivity to sources larger than about $5''$
in angular size.  The field center of the radio observations was at
R.A.=$13^h12^m17.4^s$, DEC=$+42^\circ38'05.0''$ (J2000), and the
observations were taken at two frequencies, 1.355 GHz and 1.455 GHz,
each with dual circular polarization.  In order to minimize the image
distortion and loss of sensitivity over the area of sky covered by the
optical images, the data in each frequency/polarization channel were
split into 7 channels, each of 3.125-MHz bandwidth, and the data were
sampled every 5 seconds (see \S 2.4).  A few percent of the data from
each observing session were deleted for occasional short periods of
interference, during periods of known technical problems, for
telescope shadowing, and during inclement weather conditions.

   Each observing session lasted between 8 and 13 hours, with
observations alternating between 12 min on SSA13 and 3 min on the
nearby calibrator source J1327+4326, located at
$\alpha=13^h27^m20.9790^s$, $\delta=43^\circ 26' 27.997''$; J2000,
\citep{bea02}. This absolute position of J1327+4326 with respect to
the International Celestial Reference Frame has been measured with the
Very Long Baseline Array to an accuracy less than $0.01''$ \citep{ma98}.
The relative positional accuracy obtained between J1327+4326 and the
SSA13 radio field depends on their angular separation, the apriori
calibration of VLA astrometric parameters, and weather conditions.
Based on many astrometric VLA observations, the relative positional
accuracy at 1.4 GHz in the A-configuration is $<0.1''$ and we have
used this value as the minimum position error in each coordinate for
any source in the SSA13.  The amplitude scale was derived from two
observations of 3C286 on each day, assuming a flux density of 14.55 Jy
at 1.4 GHz.  The derived flux density of J1327+4326 was 0.558 Jy and
did not vary by more than 1\% over the entire observation period.  Two
weak radio sources of 0.01 Jy located about $10'$ from J1327+4326 were
noted, but their large distance from the main radio source
decorrelated their effect; we determined the overall gain calibration
with an accuracy better than 1\%.

   The telescope-based temporal gain and phase fluctuations were
determined every 15 minutes from the calibrator observations.  The
gain variations between successive calibrator scans was generally less
than 2\%, apart from the decreasing loss of sensitivity at elevations
below $15^\circ$ of about 30\% caused by the additional noise from
ground emission.  The phase fluctuations between calibrator scans were
typically $2^\circ$ to $10^\circ$ on the short to the long spacings,
respectively.  During periods of inclement weather and at sunrise or
sunset when ionospheric activity can be large, phase changes of
$15^\circ$ over a few minutes occurred on baselines longer
than about 10 km.  Linear interpolation of the gain and phase
calibration, determined from the calibrator every 15 minutes, was
applied to the SSA13 data.

\subsection {Self-Calibration and Imaging}

   The self-calibration and imaging of the VLA data were complicated
by two factors: First, sources more than about $5'$ from the field
center are distorted by the finite bandwidth of each 3 MHz channel
(chromatic aberration), the data sampling interval, and sky curvature.
Secondly, the non-circularity of the VLA primary beam produces an
apparent variability of sources because of the relative rotation
between the plane of the sky and the primary antenna sensitivity
pattern during an observation day.  Both of these effects produce
artifacts near bright sources near the $1\%$ level.  Since the
brightest source in the image at 1.4 GHz was 20 mJy, these artifacts
were about $50~\mu$Jy/beam, well above the rms noise level of
$5~\mu$Jy/beam after 91 hours of integration.  Thus, special imaging
and self-calibration techniques were necessary to reduce these
artifacts below the level expected from receiver noise alone.

After applying the calibrations, based on the measurements of
J1327+4326, a clean (deconvolved) image of radius $40'$ centered on
SSA13 was made using the AIPS task IMAGR \citep{gre88}.  We used a
data weighting scheme (called robust=0) which weights the visibility
data in order to produce a relatively smooth distribution of the
sampled synthesized aperture, and thus a point-spread function with
lower side-lobe levels than for natural weighting that considers only
the apriori rms noise per data point.  The image resolution associated
with this weighting was $1.8''$, and a $0.5''$ pixel size was used.
The SSA13 field was covered with 38 overlapping $1024\times
1024~(512''\times 512'')$ images.  The individual image size was
chosen to limit the amplitude distortion, caused by the sky curvature,
to under $5\%$ at the edge of each image.  We also made smaller images
around the 15 brightest sources within $90'$ of the field center; these
were all stronger than 1 mJy, the strongest being 20 mJy.  Because of the
relatively accurate calibration of the data using J1327+4326, the
initial radio images were of good quality, although the above
mentioned artifacts remained.

     We continued with the next calibration step, the standard
self-calibration technique (CALIB and IMAGR in AIPS), to improve the
calibration of the SSA13 data.  Because of the redundancy of the
number of baselines (351) compared with the number of telescopes (27),
it is possible to improve the telescope calibration of SSA13 using a
source model, which is in fact the first-pass clean image
\citep{cor89}.  With the self-calibration algorithm, each telescope
residual phase was determined every four minutes for each polarization
and frequency.  The signal-to-noise in each of these intervals was
about 30:1, so robust solutions were obtained.  As expected, these
residual phase determinations were generally less than $5^\circ$,
although there were times when the phases were as large as $30^\circ$;
these coincided with periods when there were significant phase change
in the original calibration with J1327+4236\footnote{The residual
amplitude calibration was also determined; however, it was close to
unity and significant departures were indicative of periods of poor
data quality, rather than improved gain calibration.}. Because these
residual phases were primarily associated with the troposphere and
ionosphere above each telescope, the four phase
determinations---derived from two frequencies and two
polarization---agreed to within a few degrees.  During the rare
periods when there was either poor agreement among the four
determinations, or lack of reasonable continuity among the telescope
solutions, further inspection indicated data problems and these data
were deleted.  With the improved phase calibration and additional data
editing, new cleaned images were then made.

     In order to remove the artifacts due to the apparent source
variability from the non-circular VLA beam shape, we next imaged and
cleaned the data in four-hour observing blocks, which produced about
three images per observing day and 24 images for the entire
experiment.  The sum of these images reproduced the cleaned image made
from all of the data, but the brightest 15 sources clearly appeared to
vary over the day.  We then subtracted the contribution from these 15
bright sources in each 4-hour time segment directly from the data
base, and formed a residual data base from which these 15
apparently-varying sources were removed.  We then imaged and deeply
cleaned this residual data in the 38 overlapping 1024x1024 images to
the $10~\mu$Jy/beam level (2.8-$\sigma$) to produce the final image.
Finally, the cleaned images of the 15 subtracted sources (averaged
over all 24 separate time-sliced images) were added to the final image
so that all of the emission from all sources is included in the final
image.

\subsection {The 1.4 GHz Image and the Initial Radio Source List}

   The final cleaned image, covering a $34'\times 34'$ region with
$1.8''$-resolution at the center, is shown in Fig.\ 1.  The image rms
noise is $4.82~\mu$Jy/beam and is uniform to 2\% over the entire
image, except very close to the six brightest sources, and within $2'$
of the central region where the rms is $4.95~\mu$Jy/beam.  The
conversion of the image to the sky sensitivity and resolution is
discussed in the next section.

We also made an image with $6''$ resolution by convolving the data.
This convolved image has an rms noise of $13.8~\mu$Jy/beam, and was
used to determine the structure and flux density of the more extended
sources.  The formal completeness level of the uncorrected image is a
peak image flux density of $25.8~\mu$Jy/beam (or $>70~\mu$Jy/beam on
the lower resolution image), which is 5.35 times the rms noise level.
Based on the almost Gaussian noise characteristics of the image, there
is less than a $0.5$\% chance that an apparent source in the image at
the completeness level is merely a noise fluctuation.  The largest
negative peak on the image of $-24.5~\mu$Jy/beam is consistent with
our chosen completeness level chosen.

   An initial list of about 900 radio sources was generated from the
region within $17'$ of the field center.  We included sources with
peak flux density greater than $\sim 20~\mu$Jy/beam on the $1.8''$-image,
or above $65~\mu$Jy/beam on the $6''$ image.  These search limits were
selected to be somewhat below the completeness level of the image.

The properties of each source in the radio image were determined by
fitting the brightness distribution to one or two elliptical Gaussian
components.  The six independent parameters for each component were
the peak flux density, the x- and y-coordinates, the major axis, the
minor axis sizes, and orientation of the major axis.  If the source
was not appreciably resolved (the difference between the peak flux
density and the integrated flux density was less than two times the
rms noise in the image), then the average of the measured peak and
integrated flux densities was assigned as the final integrated flux
density, with the appropriate diameter limits.  Another estimate of
the integrated flux density was obtained from the peak flux density on
the $6''$ resolution image, since it is less affected by resolution
degradation or instrumental distortion effects.  This estimate was used
if it was more the 2-$\sigma$ higher than that on the higher resolution
image.

For sources with an image peak flux density less than $50~\mu$Jy/beam
(10 times the rms noise), non-linear interaction of the image noise
with the deconvolution process make the integrated flux density and
the angular size parameters and error estimates somewhat more
difficult to estimate than the peak flux density and source position
\citep{win84}.  From simulations of the model fitting of weak sources
on a radio image, we determined conservative estimates of the true
angular size of the source or upper limits as a function of
signal-to-noise ratio (SNR), and the relationship of these parameters
to the integrated flux density and its estimated error.  For this
reason, the ratio of the integrated flux density to its estimated
error is often less than the ratio of the peak flux density to its
estimated error (the SNR of the peak).

\subsection {Image Correction to the Sky}

In order to obtain a true surface brightness model of a source from
the image model, three effects must be removed.  First, because of the
finite primary beam size of the individual VLA antennas the sky
sensitivity of the observations decreases with radial distance from
the field center, with a full-width at half-power (FWHP) of $31.5'$.
The radial attenuation function, which has been measured to an
accuracy of 2\%, is nearly circular, and is shown in Fig.~2 as the
primary beam sensitivity.  Second, because of the chromatic
aberration, there is an additional smearing in the image which
broadens a source in the radial direction.  For example, for a point
source that is $17'$ from the field center its peak flux density
is 62\% of its field center value although the integrated
flux density remains unchanged.  This decrease of the peak flux
density as a function of distance from the field center is also
plotted in Fig.~2.  Finally, there is also a slight decrease in
sensitivity caused by the 5-sec sampling time and the effective
sensitivity loss for this is also plotted in Fig.~2 as the time smear.
The combined attenuation of a source on the image as a function of
distance from the field center is shown by the solid curve in Fig.~2.
At a distance of $17'$ from the field center, the sky sensitivity is
24\% of that at the field center.  The chromatic aberration also
produces a decrease of the effective sky resolution in the radial
direction from the field center.  The resolution ranges from $1.8''$
at the field center to $2.9''$ at a distance of $17'$ from the field
center.

Although a radio catalog could have been made after the above analyses
and corrections to the radio image, the relatively high density of
discrete, but often extended, radio sources caused considerable
ambiguity in determining whether closely spaced radio sources should
be considered as components of {\it one independent source} or as two
independent sources.  Since radio emission at the micro-Jansky level
is often identified with galaxies between 17-mag and 24-mag, we
have established the catalog of sources using both radio and optical
information.  Hence, each cataloged radio source (sometimes composed
of several components) is associated with one optical object, or a
close group of objects.  Whatever segregation method is used to define
a cataloged source, there will be ambiguous situations, and these are
noted in \S 5.4.5.  These sources may turn out to be among the more
interesting objects in the image.
   
\section {The Optical Data}

\subsection {The Observations and Reductions}

        Observations of the SSA13 field were made with the 8-m Subaru
Telescope and the Suprime-Cam prime focus camera with a $34'\times
27'$ field of view \citep{miy98}.  Two filters were used: The z-filter
band pass is between 8600 and 9800 \AA, and the r-filter band pass is
between 5800 and 6800 \AA.  We will designate these two wavebands as
the z-band and r-band.  Data were collected on 2001 February 2, April
22, and April 23, and all nights except April 22 were photometric.
During the February run, the Suprime-Cam was still in a commissioning
phase, and 9 of the 10 chips were usable.  The missing chip producing
a $7'\times 14.5'$ hole in the SE corner of the image. During the
April observations, the camera had one chip in the NE corner with a
bad anti-reflective coating, and this reduced the throughput in this
region by $50\%$ or more.  The camera was also severely vignetted near
the edge of the field, reducing throughput by up to 60\% for objects
more than $14'$ from the center of the field.

        On 2001 February 2, we collected 1920 sec of r-band data
with the commissioning camera.  A further 2400 sec of r-band and 3230
sec of z-band data were collected on April 22 and 23.  The camera was
offset by one arc minute between exposures in a five-point dither
pattern.  The dither pattern and large steps were chosen to fill in
gaps between chips, optimize flat fielding, and allow for chip to chip
photometric calibration.  The camera was rotated by 90 degrees between
dither patterns for z-band observations to remove bleeding from bright
stars, and to move the bad chip to the NW corner for some of the
exposures.

        These data were reduced using Kaiser's IMCAT
software\footnote{http://www.ifa.hawaii.edu/$\sim$kaiser/imcat} which is
optimized for reducing large data sets on parallel computers.  Median
sky flats were used to flat field the images.  The z-band flat was
smoothed by 32 pixels to remove fringing.  A median fringe frame was
generated for the z-band images, scaled to the sky level in each
image, and subtracted.  A second phase of flattening was performed on
each set of 5 images to remove flat fielding errors caused by flexure
in the instrument.  A second order polynomial was fit to the sky level
and subtracted from each frame to remove the sky emission.  Objects
detected at 3-$\sigma$ and pixels brighter than 200 counts were then
masked out and a surface tessellated on a 64 pixel grid.  This
background was then subtracted to remove scattered light in the image.
This was particularly important in this field because of the 10th
magnitude star in the field, and scattered light from the moon in the
z-band images.  Chip to chip and exposure to exposure photometric
offsets were then calculated and an airmass correction was applied
using a standard Mauna Kea extinction curve.  The preliminary
photometric zero points were calculated using the HDF-N as a standard
frame.  Landolt standards were also taken, but were found to be less
reliable because of the short exposure times required to ensure that they did
not saturate.  The four-minute readout of Suprime-Cam during these
runs also limited the number of standard star exposures which could be
taken.  Further calibration of the magnitude scale is described in \S
3.4.

       The Subaru images were then re-sampled onto a final grid of
$0.2''$ pixel-separation and a photometric distortion correction was
applied.  The re-sampled images were then median combined, weighted by
the rms noise in each chip and in each image.  The final images have a
usable field of view of $34'\times 27'$ in r-band and $34'\times 34'$
in z-band.  The resolution of both images, determined from the minimum
angular size of stellar-like objects, is $1.1"$.  The 3-$\sigma$
limiting magnitude is 26.1 in r-band and 24.9 in z-band, and was
determined by laying down blank $3''$ apertures on the image at least
$15''$ from a detected object.

\subsection{The Optical Images}

    The z-band (9200\AA) Subaru optical image is shown in Fig.\ 3.  The
detection level in the image is 24.9 mag (3-$\sigma$) and objects
brighter than a peak magnitude of 18 to 19 mag-sec$^{-2}$ become
saturated.  There are some evident artifacts associated with the bright 10-mag
star, and there are other blemishes near the southern edge of the
image.

\subsection {Final Radio/Optical Registration}
 
Objects brighter than 19-mag are saturated in the r-band images,
making it difficult to reliably register the images to the USNO-A
catalog (http://archive.eso.org/servers/usnoa).  To avoid this
problem, an initial astrometric solution was calculated using a
standard star field.  This resolution was applied to each individual
solution, and these corrected images were then registered to one another.
A refined astrometric distortion was then calculated for each image by
minimizing the scatter in the position of stars from image to image.
In the absence of an external astrometric reference, several
systematics are present.  First, anamorphic magnification from
differential atmospheric refraction is present in the images.
Secondly, in addition to the readout problem in one of the
commissioning chips, the relative position of the chips was misaligned
by up to $1''$, especially in declination.

   The final detailed registration of the r-band and z-band optical
images to the radio image was determined by comparing the positions of
bright, small-diameter radio sources with an obvious optical
counterpart in order to determine the small offset, rotation and
magnification between the radio and optical images, and the
chip-to-chip registration offsets.  There are over 400 secure
identifications, but about 25\% of the identifications have significant
radio-optical offsets, and 50\% are not of high
astrometric quality since they are associated with distorted galaxies
and/or extended and complex radio structures.  Hence, a total of 95
high-quality identifications established the final radio/optical
alignment.  The distribution of the radio-optical positions after the
registration is shown in Fig.~4.  The dashed circle with radius of
$0.2''$ shows the one-sigma registration error.  Since the radio
source and galaxy positions have a typical positional accuracy of
about $0.1''$, the scatter is consistent with the measurement errors
with perhaps small residual radio-optical offsets.  The quality of the
registration clearly suggests that measured offsets between the radio
and optical emission peaks larger than $0.5''$ are real offsets
associated with the source, at least for sources well above the radio
and optical thresholds.

\subsection {Magnitude Calibration and Measurement}

The final calibration of the zero point of the magnitude scale was
determined by selecting over a 100 isolated, compact optical objects,
securely identified with a radio source, that were also in the Sloan
Digital Sky Survey
(SDSS)\footnote{http://www.sdss.org/science/index.html}, then
comparing SDSS and our Subaru magnitudes.  Fig.~5 shows the linearity
and scatter of the Subaru magnitudes of isolated sources in the SSA13
catalog, listed in Table 2, versus the SDSS magnitudes (interpolated
to r- and z-bands).  After this calibration, the rms error per galaxy
is 0.3-mag in r-band and 0.5-mag in z-band, but there is no systematic
difference greater than 0.2-mag.
 
   We then determined the r-band and z-band magnitude of all relevant
objects at or near each radio source, using this SDSS-based magnitude
calibration.  For objects fainter than about 22 mag, the magnitude was
determined by the sum of intensities within a radius of $3''$ from the
galaxy center.  For brighter and more extended galaxies the area was
increased as necessary.  For close binary galaxy-pairs, the area was
adjusted to include only the emission from the relevant galaxy.  The
emission associated with faint halos of galaxies was not included in
the estimated magnitudes.  In all cases, the same area was used to
determine the magnitude for both filters.  For radio sources with no
apparent optical counterpart, we used the 2-$\sigma$ noise level as the
magnitude limit.  This limit increased near the image edges and near
the bright star in the field.

\section {The Radio/Optical Catalog}

\subsection {Compilation of the Radio/Optical Catalog}

   We then carefully compared the optical and radio fields for the
preliminary list of 900 radio sources which we had compiled using only
the radio data.  With the radio and optical registration of $0.2''$,
objects as faint as z=24.9 and r=26.2 could be identified with a radio
source, based on proximity alone.  Thus, the astrometric accuracy was
crucial in determining the most probable identification or lack of
identification for the radio sources, some of which are extended by
several arcsec and are located in relatively crowded optical fields.
It is also clear that the centroids of some radio sources are
significantly offset from the center of the optical counterpart.  The
individual identifications are discussed more fully in \S 4.5.

\subsection {The Complete Sample}

    The radio/optical catalog, listed in Table 2, contains 810 radio
sources within $17'$ of the field center.  Because of the significant
attenuation and distortion of sources near the edge of the
field, we limited the area of formal completeness to a radius of
$15'$ from the field center, and a peak {\it image} flux density
greater than $25.8~\mu$Jy/beam.  This corresponds to a {\it sky}
completeness of $75~\mu$Jy/beam within the $15'$ radius.

     Of the 810 sources in the catalog, 560 sources are in this
complete sample, and it is only this sample which is used for
statistical analysis of the properties of the radio sources.  The
additional 250 sources, while not part of the complete sample, are
useful additions in understanding the complex radio and optical
morphologies found at the lower flux density levels.  The probability
that these weak sources are real is $>95$\%.  Thus we estimate that
fewer than ten to fifteen of the 810 cataloged sources may not be
real.

\subsection {The SSA13 Radio/Optical Catalog}

    The radio/optical catalog of the SSA13 Field is given in Table 2.
It is arranged in right ascension order, although components of a
single source are kept together.

\begin{itemize}
\item {} Column 1: The radio source number used in this paper.  The
letter {\bf N} before the number indicates that it is {\bf NOT} in the
complete sample: either the peak flux density on the image is less
than $25.8~\mu$Jy/beam; or it lies more than $15'$ from the field center.
If the radio source is composed of several discrete radio components,
these are separately listed underneath as A, B, etc.
\item {} Column 2: The IAU designation for each source.  Individual
radio components of a cataloged source are not given separate IAU
designations, but should use that of the source as a whole.
\item {} Column 3: The SNR of the radio source
and radio components, given by the peak
image flux density divided by the rms noise.
Those sources with SNR $<5.35$ are not in the
complete catalog.
\item {} Column 4: The source integrated flux density and error
estimates in microJansky,  {\it corrected to the sky for all
instrumental effects}.  The determination of this parameter is discussed
in \S 2.4.
\item {} Column 5: The right ascension and error estimate of the radio
emission.  For sources larger than $8''$, the position is that of the
brightest emission peak in the source, not the centroid.  The minimum
error is 0.010s, as discussed in \S 2.1.
\item {} Column 6: The declination and error estimate defined as in
column 5.  The minimum error is $0.10''$.
\item {} Columns 7, 8, 9: The radio FWHP major axis and minor axis in
arcsec, and position angle, north through east, in degrees.  All
instrumental and noise corrections have been made and this angular
size is the best estimate of the true angular size or its limit.  The
size of each individual radio component is also given.  The angular
size limit is a strong function of SNR.  Based on simulations, we
estimate that there is more than $85\%$ confidence that a source with
a tabulated angular size is really resolved.
\item {} Column 10: The radio/optical morphology category (ROM).
These are defined in Table 3 and discussed in the next section.  The
major categories are: EG = Extended Active Galaxies; S = Stellar
Object; G = Galaxy; B = Binary Galaxy System; C = Complex; F = Faint
ID; U=Unidentified.  The sub-categories in lower case letter are: c =
radio emission centered on optical object; e = extended radio source;
d = radio source is displaced from optical center. There is some
subjectivity among the categories.  Individual optical identifications
for radio components in a cataloged source are also listed when
applicable.  A question mark ({\bf ?})  indicates that there is a note
associated with this catalog entry.  For the small portion of the
radio field covered only by the SDSS and not the deeper Subaru images,
the optical classifications are limited to: sg=galaxy,
su=unidentified.
\item {} Column 11: The r-band integrated magnitude of the optical
counterpart to the radio source (or components where applicable).
\item {} Column 12: The z-band integrated magnitude of the optical
counterpart.
\end{itemize}

\subsection {The SSA13 Radio/Optical Plots}

The comparison of the radio and optical emission for all 810 sources
in the catalog is shown in Figs.~6-1 to 6-33.  Most fields in these
figures are $12''\times 12''$ in size, although some plots are as
large $30''\times 30''$ to display all of the radio and optical
emission of extended or complex sources. {\it The full-set of plots
are found in
ftp://ftp.cv.nrao.edu:/pub/NRAO-staff/efomalon/ssa13\_diagrams/.}

The contours show the {\it radio} emission with the lowest contour
value at $10.0~\mu$Jy/beam, and levels at -1,1,2,4,16 times this
value.  For the brightest radio sources, the lowest radio contour
level has been increased to remove low level artifacts.  The colored
display shows the optical emission, usually at z-band unless the
r-band image is of superior quality.  The color transfer function for
each source has been adjusted in order to show the salient
morphological features of the optical emission, and to limit the noise
contribution in regions of lower optical sensitivity.  For those
sources with no optical information from Subaru, only the radio
contours are shown, with a pink color background; however, Table 2
does contain any identifications and magnitudes obtained from the
SDSS.

\subsection {Comments on Individual Sources}

   Individual comments on selected radio/optical fields are given
below.  Most references to SDSS are associated with point-like optical
emission where the colors can distinguish between stars and quasars.
                                                                        
\footnotesize                                                                  
\vskip 0.3cm\noindent                                                   
009.~~The smaller object to the south-west is a star (SDSS).
\vskip 0.3cm\noindent                                                   
020.~~The smaller object to the west is a star (SDSS).
\vskip 0.3cm\noindent                                                   
021.~~The radio emission is displaced north-west of the optical center.
It is extended to the east and is associated with a faint
optical galaxy.
\vskip 0.0cm\noindent                                                   
027.~~The extended radio source is about $8''$ south of group of
bright galaxies.  It is unidentified, but faint background optical
emission lies near the radio emission.
\vskip 0.0cm\noindent                                                   
032.~~The radio source is well displaced from the galactic nucleus.     
There is a spur of optical emission near the radio centroid.      
\vskip 0.0cm\noindent                                                   
050.~~The main part of the radio emission is associated with a 23-mag
galaxy.  There is radio emission extending to the galaxies to the east
and west.
\vskip 0.0cm\noindent                                                   
053.~~The extended radio emission is associated with a close binary
system.
\vskip 0.0cm\noindent                                                   
069.~~This extended radio source is identified with a symmetric
galaxy.  The contour plot of the entire radio source (Fig.~8) shows a
morphology closer to FRI than FRII (lobes are diffuse with no hot spot).
\vskip 0.0cm\noindent                                                   
091.~~The radio source is displaced $3''$ west of the galaxy nucleus.
It may be a background unidentified source, or it may be
associated with a faint spur toward the west from the galaxy.
\vskip 0.0cm\noindent                                                   
096.~~The radio source is identified with a faint galaxy.  There is
extended radio emission connected to another galaxy about $7''$ to the
south.
\vskip 0.0cm\noindent                                                   
101.~~One radio component is associated with the nucleus of the
galaxy.  A second radio component (101B) is located $2''$ north,
roughly in the direction of the extended optical core.
\vskip 0.0cm\noindent                                                   
104.~~The radio source is located at the nucleus of the southern
galaxy of a close pair.  Source 105 is to the north.  The circular
object to the west is a star (SDSS).
\vskip 0.0cm\noindent                                                   
105.~~The radio source is located at the nucleus of the northern
galaxy of a close pair.  Source 104 is to the south.
\vskip 0.0cm\noindent                                                   
109.~~Most of the radio emission is associated with the southern
galaxy.  The object about $3''$ to the north-west is a star (SDSS),
with a hint of faint radio emission.
\vskip 0.0cm\noindent                                                   
112.~~The object $4''$ to the west is a star (SDSS).
\vskip 0.0cm\noindent                                                   
122.~~The radio source lies in the middle of a group of galaxies,
but is considered unidentified since there is no optical emission
coincident with the radio source.
\vskip 0.0cm\noindent                                                   
141.~~The identification is uncertain.
\vskip 0.0cm\noindent                                                   
143.~~The main radio emission is associated with a 23-mag galaxy.
There are indications of extended emission associated with two
nearby galaxies.
\vskip 0.0cm\noindent                                                   
156.~~The radio source is located near a bright star (SDSS)
that is saturated on the image.  Source 158 is on the left
lower border of the frame.
\vskip 0.0cm\noindent                                                   
158.~~The radio emission is located within the glare of
a bright star (SDSS).
\vskip 0.0cm\noindent                                                   
159.~~The identification is uncertain.
\vskip 0.0cm\noindent                                                   
169.~~The radio source is associated with a blue stellar
object (SDSS).  It could be a quasar.
\vskip 0.0cm\noindent                                                   
171.~~The identification is uncertain.
\vskip 0.0cm\noindent                                                   
175.~~The large-scale radio emission is associated with several
galaxies.
\vskip 0.0cm\noindent                                                   
177.~~The identification is uncertain.
\vskip 0.0cm\noindent                                                   
183.~~The optical object $3''$ to the west is a star (SDSS).
\vskip 0.0cm\noindent 
185.~~Three objects show radio emission.
However, the northern-most radio source (185C) is a slightly red stellar
object (SDSS), while components 185A and 185B are associated with individual
galaxies.
\vskip 0.0cm\noindent                                                   
188.~~The radio emission associated with the elliptical galaxy is
complex.  A bright star is located about $10''$ north-east.
\vskip 0.0cm\noindent 
195.~~The radio emission lies between a circular
galaxy to the north and a fainter galaxy to the south-west (morphology
unknown).
\vskip 0.0cm\noindent 201.~~Two of the three galaxies have radio
emission (components 201A and 201B).  They are cataloged as one
radio source since the galaxies appear interacting.
\vskip 0.0cm\noindent                                                   
212.~~The radio contours here just show the radio core of an extended
active galaxy with lobes $38''$ to the south and $45''$ to the north.  See the
contour plot of the entire source in Fig.~8.  This source has a classic
FRII morphology.
\vskip 0.0cm\noindent                                                   
229.~~The radio source is lost in the glare of a bright star (SDSS).
\vskip 0.0cm\noindent                                                   
234.~~The identification is uncertain.
\vskip 0.0cm\noindent                                                   
236.~~The bright object to the north-east is a star (SDSS).
\vskip 0.0cm\noindent                                                   
239.~~The fainter object to the south-west is a star (SDSS).
\vskip 0.0cm\noindent                                                   
244.~~The brighter radio source is identified with a 20-mag, slightly
red stellar object (SDSS).  The weaker component $3''$ to the
south-west is unidentified.  Perhaps it is a background galaxy which
is obscured.
\vskip 0.0cm\noindent 
254.~~Unidentified.  With source 255 it could be
a pair of radio lobes associated with the galaxy located between them.
\vskip 0.0cm\noindent                                                   
255.~~See note for source 254.
\vskip 0.0cm\noindent                                                   
257.~~Extended radio emission has a similar orientation to the           
optical galaxy.  This source is $10''$ west of 263.                           
\vskip 0.0cm\noindent 
263.~~The orientation of the radio emission is similar
to the optical light of the galaxy, and points to a fainter binary
component.  This source is $10''$ east of 257, and is assumed to be an
independent source.
\vskip 0.0cm\noindent                                                   
264.~~Most of the radio emission is offset from the nucleus of the
galaxy, but coincident with a faint optical feature.  It is classified
as a binary.
\vskip 0.0cm\noindent                                                   
266.~~There is extended radio emission along the minor axis of the
galaxy.  It could be an extended AGN.
\vskip 0.0cm\noindent                                                   
267.~~The identification is uncertain.
\vskip 0.0cm\noindent                                                   
268.~~The main radio component (268A) is near the galaxy center.  Two
fainter arms of radio emission are also present.
\vskip 0.0cm\noindent                                                   
275.~~The identification is uncertain.
\vskip 0.0cm\noindent                                                   
280.~~The major component is offset $2''$ south-west of a
galaxy.  A brighter galaxy with faint emission lies about $7''$ to the
north.
\vskip 0.0cm\noindent                                                   
294.~~The relationship of the radio emission and the faint
galaxies within $2''$ is unclear.
\vskip 0.0cm\noindent                                                   
295.~~The small-diameter radio source is significantly displaced
from the center of an elliptical galaxy.  There is a star (SDSS)
located $8''$ to the north-east.
\vskip 0.0cm\noindent
299.~~The optical object is a red stellar object (SDSS).
\vskip 0.0cm\noindent 304.~~An extended active galaxy with an
intermediate FRI-II morphology.  See contour plot in Fig.~8.
\vskip 0.0cm\noindent                                                   
314.~~Most of the extended radio emission is located in the center of
a bright galaxy; however there is extended radio emission toward
a faint companion to the west.
\vskip 0.0cm\noindent 
315.~~A group of galaxies contains radio
emission associated with two of the galaxies.  There is additional
extended radio emission.
\vskip 0.0cm\noindent                                                   
319.~~The radio emission is displaced south-west of the core of the
galaxy.  The relationship to the bright galaxy $8''$ to the north-east
is unclear.
\vskip 0.0cm\noindent                                                   
330.~~A complicated group of galaxies.  Two galaxies have            
significant radio emission (330A, 330B).  The source 335 is
south-east of this region.                             
\vskip 0.0cm\noindent                                                   
335.~~A line of three galaxies, with the southern two having radio
emission.  The radio sources are $15''$ south-east of the complex of
sources 330.
\vskip 0.0cm\noindent                                                   
340.~~The radio source is in the glare of the 10-mag star.
\vskip 0.0cm\noindent
346.~~The radio emission contains two
components, each probably identified with a faint galaxy.  It is
classified as a binary, but could be considered as two independent
identifications.
\vskip 0.0cm\noindent                                                   
350.~~The radio source is close to the 10-mag star.
\vskip 0.0cm\noindent                                                   
353.~~The radio emission is relatively weak and extended.
It is cataloged as one source with two
components, the southern one identified with a galaxy and the more extended
northern component leading toward a fainter galaxy to the north-east.
However, the source is weak and this association is uncertain.
\vskip 0.0cm\noindent
359.~~The radio source is identified with a faint galaxy.  There is an
elliptical galaxy $2''$ to the west, and it is unknown if there is a
physical connection between these two optical objects.
\vskip 0.0cm\noindent
360.~~The radio source is too close to a 10-mag star to be identified.
\vskip 0.0cm\noindent                                                   
385.~~The two radio components are identified with two galaxies in a
crowded group of galaxies.  The bright object to the south-east is
stellar.
\vskip 0.0cm\noindent
386.~~The radio source is unidentified.  Source
391, which is $6''$ north-west is also unidentified.  It is possible
that 386 and 391 are radio lobes of the galaxy lying roughly between
them, but there is no detection of a radio core.
\vskip 0.0cm\noindent
390.~~The object north of the radio identification is stellar (SDSS).
\vskip 0.0cm\noindent
391.~~The radio source is unidentified.  Source 386, which is $6''$
south-east, is also unidentified.  It is possible that 386 and 391 are
radio lobes associated with the galaxy between them.
\vskip 0.0cm\noindent                                                   
400.~~The redshift is 0.322 \citep{fom02}.
\vskip 0.0cm\noindent                                                   
403.~~The optical object is blue stellar object (SDSS).  It is a quasar
with a redshift of 2.561 \citep{fom02}.
\vskip 0.0cm\noindent                                                   
407.~~The radio source is associated with a faint, diffuse object that
is $2''$ east of a brighter galaxy.  The two galaxies may be part of a
binary.
\vskip 0.0cm\noindent 
409.~~The peak emission of the radio source lies
$1.5''$ south-west of the center of the galaxy, and may be associated
with an optical spur to the south-west.
\vskip 0.0cm\noindent                                                   
412.~~The optical object is a red stellar object (SDSS).
\vskip 0.0cm\noindent                                                   
413.~~The radio emission is associated with a galaxy.  The bright
object is stellar (SDSS).  The low surface brightness of the extended
optical emission is puzzling.
\vskip 0.0cm\noindent                                                   
416.~~The optical object is a blue stellar object (SDSS).  It is
probably a quasar.
\vskip 0.0cm\noindent                                                   
418.~~The extended radio source is identified with a faint object at
the edge of the image of a bright star (SDSS).
\vskip 0.0cm\noindent                                                   
419.~~The optical object is a blue stellar object (SDSS).  It is
probably a quasar.
\vskip 0.0cm\noindent 
431.~~The nature of this radio source is puzzling.  One radio
component (B) is associated with faint emission south-east of the
nucleus of the brighter object.  The brighter radio source is
north-west of the bright galaxy, although it could be associated with
a faint spur (or an unrelated faint background object) to the
north-west.  These components could also be the radio lobes associated
with the bright galaxy with no detectable core emission from the
nucleus.  However, the near coincidence of this possible radio lobe to
the south-east with optical emission argues somewhat against a
twin-lobed radio source.  The 8-GHz radio image (source
29=J131219+423608 in the 8-GHz catalog) with lower resolution blends
these two radio
components with no indication of emission from the galaxy nucleus
\citep{fom02}.
\vskip 0.0cm\noindent                                                   
434.~~The identification is uncertain.
\vskip 0.0cm\noindent                                                   
435.~~A complex region of radio and optical emission.  The peak of the
radio emission is near a faint spur from the most southern galaxy.
\vskip 0.0cm\noindent                                                   
438.~~The redshift is 0.302 \citep{fom02}.
\vskip 0.0cm\noindent                                                   
442.~~The redshift is 0.180 \citep{fom02}.
\vskip 0.0cm\noindent                                                   
443.~~The bright radio component lies near the center of the
galaxy.  There is also a faint radio component (443B)
$4''$ to the east.  It is unknown if this is related to the galaxy or
associated with a faint background object.
\vskip 0.0cm\noindent 
446.~~The peak of the radio emission is
coincident with a circular galaxy.  Radio jets emanate north and south
of the core for $5''$, and the source has been classified as an
FRI galaxy.  See contour plot in Fig.~8.
\vskip 0.0cm\noindent                                                   
449.~~The optical object north-east of the radio source is stellar
 (SDSS).  The identification in uncertain.
\vskip 0.0cm\noindent 
450.~~The radio source is centered on a blue
stellar object (SDSS), a quasar with redshift 2.561 \citep{fom02}.
\vskip 0.0cm\noindent                                                   
453.~~The object north-east of the radio source is a red stellar
object (SDSS).  Hence, it is unlikely to be part of a binary system.
\vskip 0.0cm\noindent                                                   
457.~~The redshifts of both galaxies are 0.401 \citep{fom02}.
\vskip 0.0cm\noindent                                                   
461.~~Some of the radio emission extends from the center of the
fainter galaxy toward the brighter galaxy.
\vskip 0.0cm\noindent                                                   
462.~~The radio emission is extended and may be associated with
three closely-grouped galaxies.
\vskip 0.0cm\noindent                                                   
466.~~The fainter extended source of a close, weak pair.  Source 471
about $3"$ to the north-east is more compact.  Both sources are
unidentified and not in the complete catalog.
\vskip 0.0cm\noindent                                                   
469.~~The brighter radio source lies near the center of a galaxy.  The  
fainter radio source (469B) is associated with another galaxy to the    
east.  These two objects have been cataloged as one source because the
galaxies appear to be interacting.
\vskip 0.0cm\noindent                                                   
471.~~The brighter compact source of a close, weak pair.  Source 466
about $3"$ to the south-west is more extended.  Both sources are
unidentified and not in the complete catalog.
\vskip 0.0cm\noindent                                                   
477.~~There is faint extended radio emission from the optical 
identification toward another galaxy $8''$ to the north.
\vskip 0.0cm\noindent                                                   
483.~~The radio source is identified with an elliptical galaxy that
is near a bright red stellar object (SDSS).
\vskip 0.0cm\noindent                                                   
490.~~The radio source is coincident with a galaxy that is about $2''$
south of a stellar object (SDSS).
\vskip 0.0cm\noindent 
503.~~An extended active galaxy with a morphology closer to FRI
than to FRII.  See contour plot in Fig.~8.
\vskip 0.0cm\noindent                                                   
504.~~The radio source is coincident with a a galaxy that lies near
two bright stars (SDSS).
\vskip 0.0cm\noindent                                                   
507.~~This radio source is north of the northern lobe of the extended
active galaxy 503 (Fig.~8).  It is identified with a faint galaxy and
probably not part of source 503.
\vskip 0.0cm\noindent 
515.~~The main radio source is located $1''$
east of a bright red stellar object (SDSS).  This identification is
classified as stellar, but may be a compact red galaxy.  There may be
an extended, faint radio emission along the galaxy north-south line.
\vskip 0.0cm\noindent 
535.~~The radio emission lies on a galaxy to the north of a bright star
(SDSS).
\vskip 0.0cm\noindent                                                   
549.~~The strong radio core and the emission to the west and east       
are associated with an extended active galaxy. It could be
an FRII source, but the lobes are too weak to be certain.
See Fig.~8.                
\vskip 0.0cm\noindent 
563.~~The radio source is coincident with a
faint optical object that is $5''$ east of a bright circular galaxy.
We have classified this as a binary system.
\vskip 0.0cm\noindent                                                   
566.~~The compact radio emission is located $3''$ west of a bright
galaxy, perhaps identified with a faint optical object.  A faint radio
extension toward the bright galaxy suggests that the compact radio
emission is associated with the bright galaxy.
\vskip 0.0cm\noindent                                                   
567.~~The two radio components have been classified as a binary;
however, they may be independent sources.
\vskip 0.0cm\noindent                                                   
569.  The nucleus of this bright galaxy has a slightly extended
radio source.  Source 578 is located $10''$ to the north-west.
\vskip 0.0cm\noindent                                                   
575.~~The slightly extended radio source is identified with a faint
galaxy.  The bright object to the west is stellar (SDSS).
\vskip 0.0cm\noindent                                                   
582.~~The radio source is associated with a faint binary visible above
the glare of the nearby bright star.
\vskip 0.0cm\noindent                                                   
589.~~The identification is uncertain.
\vskip 0.0cm\noindent                                                   
594.~~Most of the radio emission comes from an optical spur to the
north-east of the galaxy nucleus.  Some extended emission covers the
other spur and the galactic nucleus.
\vskip 0.0cm\noindent
601.~~The radio emission is located at the western edge of a 25-mag
galaxy which is probably in a binary pair.
\vskip 0.0cm\noindent
611.~~The radio source is centered on a faint object between two
brighter objects. 
\vskip 0.0cm\noindent                                                   
621.~~The extended radio emission is an asymmetric FRI source.
See Fig.~8.
\vskip 0.0cm\noindent                                                   
626.~~The radio source is clearly displaced from the galaxy.  The
probability that this is a random association is about 10\%.
\vskip 0.0cm\noindent
634.~~The radio emission is coincident with a galaxy that is located
in the glare of the bright star.
\vskip 0.0cm\noindent                                                   
654.~~This region, near the plate edge, has poor optical sensitivity.
The western component might be identified with a faint galaxy, the eastern
is not.  Classified as unidentified.
\vskip 0.0cm\noindent                                                   
660.~~The radio emission lies on an optical bridge between two          
galaxies.                                                               
\vskip 0.0cm\noindent                                                   
668.~~A large diffuse radio source with two extended components
is associated with a group of galaxies.
\vskip 0.0cm\noindent 676.~~The radio component is clearly displaced
from a galaxy core and not associated with an optical feature.  It is
unlikely to be associated with another, invisible background
galaxy. The probability that this source is unrelated to the bright
galaxy is $<10\%$. The source to the east is 680. 
\vskip 0.00cm\noindent                                                  
680.~~The radio source is coincident with a compact galaxy which is
$7''$ east of a brighter galaxy.  Source 676 lies to the west.
\vskip 0.0cm\noindent                                                   
685.~~The radio source is associated with a faint galaxy near the edge
of the image of a bright star (SDSS).  The identification is uncertain.
\vskip 0.0cm\noindent                                                   
711.~~The two galaxies in this binary each have radio emission which is
oriented along each galaxy's major axis.  They could be considered as
two independent sources, although the optical objects appear interacting.
\vskip 0.0cm\noindent                                                   
715.~~The identification is uncertain.
\vskip 0.0cm\noindent 
721.~~The identification with a very faint
optical object has only a 50\% probability.  Faint extended radio
emission to the east may be associated with another galaxy $3''$
to the east.
\vskip 0.0cm\noindent                                                   
738.~~The radio emission is extended and covers several
galaxies.
\vskip 0.0cm\noindent                                                   
740.~~The unresolved radio emission peak is located between two 23-mag
galaxies.  It is considered as unidentified, although it may be
associated to one or both of the nearby galaxies.  Source 741 can be
seen in the upper left and is assumed to be unrelated to 740.
\vskip 0.0cm\noindent                                                   
748.~~A bright source at the edge of the z-band image with reduced
optical sensitivity and some distortion.  The identification is uncertain.
\vskip 0.0cm\noindent                                                 
753.~~The radio emission is extended and may be associated with several
galaxies in this group.
\vskip 0.0cm\noindent                                                   
754.~~The radio source is close to a 26-mag object that is $1''$
north of a brighter galaxy.  This identification is uncertain.
\vskip 0.0cm\noindent                                                   
771.~~The radio source is extended, and located between the galaxy
nucleus and an optically brightened region about $1''$ to the east.
\vskip 0.0cm\noindent                                                   
779.~~The probability of identification of this radio source with the
faint object that is $2''$ north of a binary system is 80\%.  It is
unknown if the faint optical emission is associated with the brighter
binary.
\vskip 0.0cm\noindent                                                   
790.~~The radio source lies $1''$ north of a 24-mag galaxy.  It could
be associated with this galaxy, but may also be identified with a
faint optical object.  It is tentatively considered as unidentified.
\vskip 0.0cm\noindent 792.~~The extended radio source is coincident
with a faint galaxy that is $8''$ south-west of a much brighter
galaxy.  We assume that the faint galaxy is a background object.
\vskip 0.0cm\noindent                                                   
800.~~The radio source is associated with faint diffuse emission that
is $3''$ north of a brighter galaxy.  There is also faint radio emission
near the bright galaxy.
\vskip 0.0cm\noindent 806.~~The extended radio source is centered
on a faint object that lies between two binary systems.  The radio
emission extends to the northern galaxy of a binary system to the
north-west.
                                                                        
\normalsize               

\section {Optical Properties of Radio Sources}

\subsection {Identification Rate}

The identification rate for the sources in the complete sample, as
indicated in Table 2, is close to 90\%.  Of the 551 sources for which
there is z-band and/or r-band Subaru coverage (nine sources have only
SDSS coverage), 499 (90.6\%) are optically identified with a
3-$\sigma$ detection level of r=26.1 mag or z=24.9 mag.  In
determining an identification of a radio source, the high optical and
radio resolution and the $0.2''$-rms astrometric alignment of the
images allowed us to use optical and radio structure properties, as
well as just the radio and optical centroid offset.  Ignoring this
additional information and using straight-forward identification
algorithms to calculate likelihood percentages of identifications
\citep{sut92} that are based solely on the radio-optical offset and
the radio and optical densities gives a somewhat lower rate of
identifications.

    For a more conservative determination of the identification rate,
we have excluded faint identifications (Table 2, ROM=F?) where there
is some doubt ($<75\%$ chance) about the identification, and excluded
all identifications with a significant optical radio offset (Table 2,
ROM=Gd) although there may be morphological and statistical evidence
for an association.  The number of identifications then drops from 499
to 483 for the more conservative identification rate of 87.7\%.  This
rate is similar to the i-band (~7700\AA) identification rate of 85\% for
the HDF above a limited magnitude of 25-mag \citep{ric98,mux05} and
the 82\% k-band detection rate of the deepest Phoenix Deep Field (PDF)
\citep{sul04}.  For the brighter galaxies alone, the identification
rate from the entire PDF is 73\% down to r=24.0 mag \citep{afo05},
whereas the identification rate from the SSA13 field to the same
magnitude limit is 55\% (269 ID's out of 489).  Hence, the
identification rate is about 65\% for a limit of r=24.  We conclude
that the identification rate of microJansky radio sources in the SSA13
field is at least $88\%\pm 2\%$ with r-band and z-band magnitude limits of 26.1
and 24.9, respectively.

\subsection {Magnitude Distributions}

      The distribution of magnitudes for the complete sample of
sources are shown in Fig.~7a for z-band and Fig.~7b for r-band.  The
number distribution for the z-band galaxy magnitudes is relatively
flat between 20-mag and 23-mag (per magnitude interval), and then
drops off significantly for fainter galaxies.  Only 9\% of the radio
sources are unidentified.  The distribution of the r-band magnitudes,
on the other hand, continues to rise with a peak at about r=24.0 mag
and a drop-off at fainter magnitudes.  Since only 11\% of the radio
sources are not identified in this band, the drop-off must be real.
The drop-off in the number distribution of identifications at r=23 mag
in the PDF survey \citep{afo05} (in their Fig.~1) is consistent with
their optical detection level limit of r=24 mag, and their lower radio
resolution that makes their identifications with faint objects more
ambiguous.

    The color distribution (r-z magnitude) for the SSA13 complete
sample with detections at both bands (440 sources) is shown in
Fig.~7c.  The average r-band (6300A) minus z-band (9200A) magnitude
difference is 1.3 mag, and about 25\% of the galaxies have a reddening
(r-z) greater than $1.8$ mag.  This amount of r-z reddening
corresponds to the $r-k$ band reddening demarcation of about 5.0 mag
which defines an extremely red object (ERO).  Hence, the percentage of
microJansky sources in the SSA13 which are ERO's is consistent with
that found using the wider-separated wave bands \citep{mor00}.

The histogram also shows that the fainter galaxies near r=24 mag are
significantly redder than galaxies with r=22 mag.  This is consistent
with the correlation of redshift and identification magnitude found in
the PDF \citep{afo05}: the fainter optical sources are more redshifted
so that restframe emission is at shorter wavelengths where the
reddening is more apparent.
                                              
\section {The Radio/Optical Morphologies (ROM)}

    Two major goals of this paper are the presentation of the catalog of
microJansky radio sources in the SSA13 field, and the comparison of the
radio and optical morphologies of these sources. In order to
facilitate the concise description of the radio and optical properties
of a source, we use a radio/optical morphological (ROM)
classification system used in Column 10 of Table 2 and described in
the table headings.  The number of sources in each of the
classifications is given in Table 3, both for all 810 sources and for
those in the complete sample of 560 sources.  The classification of
single and binary systems, and the reliability of the faint
identifications, are discussed below.

\subsection {Radio Sources Associated with  Isolated Galaxies: ROM=G}

Of the 560 source in the complete sample, only 390 had identifications
which are sufficiently bright so that an assessment of the optical
complexity could be determined.  About 60\% of these sources have
an optical identifications (z$<24.0$) to allow us to classify them as
associated with a relatively isolated galaxy.  We use the average
density of galaxies from SDSS in the SSA13 field and the HDF down to
25 mag, to determine the average separation of galaxies brighter than
r=25m, 24m, 23m, 22m, 21m and 20m, which is $7'', 11'', 17'', 25'',
37''$, and $55''$, respectively.  If two galaxies are closer than
about one-third the expected mean separation (corresponding to 10\% or
less chance of random associated), they are probably a physical pair.
However, other considerations related to the morphology and shape of
the neighboring galaxies and radio sources are also used to
distinguish isolated galaxies from binary or more complex systems.

   The radio emission from about 40\% of the isolated galaxies is
resolved, with an angular size that is typically $<1.2''$.  The
orientation of the radio emission is often related to the orientation
of the galaxy, as shown in the following sources: 72, 90, 135, 180,
190, 194, 196, 198, 277, 289, 307, 308, 320, 403, 430, 447, 459, 514,
543, 590, 628, 670 and 758.  In most cases the radio orientation is
aligned along the galaxy major axis, but some sources have a clear
radio extension along the minor axis.

    About 8\% (30 out of 390) of the single galaxy identifications
have radio components clearly displaced by several arcsec from
the galaxy nucleus (ROM=Gd), although some radio emission may also
be associated with the galaxy nucleus.  In many cases there is optical
enhancement or asymmetry outside of the central region that appears
associated with the radio emission; hence, they may be in binaries in
the latter stages of merging.  Some of these sources are: 15, 38, 186,
271, 345, 497, 515, 519, 532, 594, 679 and 690.  Two sources, 91 and
676, are significantly displaced from the galaxy center (still only
$3''$), but the identifications are likely, based on the probability
of $<10\%$ that these are a chance occurrence.  

\subsection {Binary and Complex Systems: ROM = Bc, Be, B2, C}

   About 40\% of the radio sources with sufficiently bright optical
counterparts are associated with binary or more complicated optical
systems (ROM=B or C).  Since most of the radio emission is
usually confined to only one galaxy, this classification is determined
by the proximity of other galaxies, and the morphology and shape of
the optical and radio emission that suggest these are interacting
systems.

   Binary systems which contain a majority of the radio emission in an
unresolved radio source in the center of one of the galaxies are
common.  Some examples are: 9, 23, 84, 109, 127, 128, 134, 140, 183,
231, 298, 341, 377, 390, 420, 508, 545, 551, 552, 643, 666, 723 and 741.
In some of these cases, there is a hint of radio emission toward or
associated with the other galaxy, but the emission is too faint to be
cataloged.  Binary systems with extended radio emission (ROM=Be) often
show a correlation between the orientation of the radio emission and
the binary position angle, and examples are: 16, 28, 49, 99, 119, 131,
199, 209, 237, 239, 279, 322, 338, 372, 392, 393, 414, 425, 453, 457,
461, 473, 502, 510, 541, 574, 579, 587, 595, 610, 639, 672, 683, 688,
707, 737, 743 and 750.  Because of the difficulty in deriving angular
sizes for the weaker sources, the physical difference between the
ROM=Bc and Be sources may be somewhat artificial.

    About 10\% of the binaries have significant radio emission
associated with both galaxies (ROM=B2) where both radio components are
strong enough to be individually cataloged.  Examples are: 35, 166,
286, 310, 361, 629, 694 and 711.  Some of the ROM=Be sources are
probably similar, but with the emission from the fainter galaxy somewhat
below our detection limit.

   Those few radio/optical sources which are labeled as complex
generally contain more than two galaxies, and/or two radio components
which are not simply related to the optical emission.  The radio
properties are often similar to those of binary systems, but with
additional complexity.  These sources are more fully discussed in the
notes to Table 2 and include numbers 185, 188, 353, 385, 431 (might be
an AGN), 435, 668, 738, 753 and 806.

\subsection {Radio Sources Associated with a Stellar Object: ROM=S}

   Eight sources are associated with stellar objects: 169, 299, 403,
412, 416, 419, 450 and 515.  Five of them---169, 403 (confirmed
redshift of 2.561), 416, 419, and 450 (confirmed redshift also of
2.561)---are blue and probably all are quasars.  Sources 299, 412 and
515 are red and probably are stars.  The radio emission of source 515
is displaced about $1''$ east of the center of the red stellar object
and there may be extended emission.  Two sources cataloged as complex
have part of their radio emission apparently associated with a stellar
object, 185 and 244.  See the notes to these sources.  Hence, up to
1\% of the microJansky sources are galactic stars, comparable to
detection rates found in other deep radio surveys.

\subsection {Extended AGNs: ROM=EA}

AGNs are defined by a radio component that is co-located with the
active nucleus of a galaxy.  However, some active galaxies also have
extended radio emission that is often bi-polar and contains narrow
jets and relatively compact radio components in the jets or in the
lobes.  These sources are called `classical doubles'. If the jets are
dominant and the hot spots are located near the beginning of
the jets, they are called FRI.  If the bright hot spots are dominant
and at the edge of the lobes, they are called FRII \citep{fan74}.  We
were able to detect these large sources because the component sizes
are generally smaller than $8''$, the angular size where our radio
observations become insensitive.  Extended AGNs with compact lobes
that extend more than about $20''$ can be difficult to identify
because the lobe may overlap with other faint radio sources.  For
example, source 212 contains a radio core that is identified with a
faint galaxy.  Two asymmetric radio sources, about $40''$ away, have
no co-located optical counterpart, have a structure that is suggestive
of an FRII double source and are symmetrically disposed with respect
to the suggested core component.

   Seven apparently extended `classical' extended radio galaxies were
found in the SSA13 field, and the corresponding radio/optical fields
are shown in Fig.~8.  Five are in the complete sample, and two are
located outside of the $15'$ region defined for the completeness
limit. All seven active galaxies are relatively strong radio sources,
with the weakest having a total flux density of $149~\mu$Jy while
three are well above 1 mJy.

    Even with this small sample, we find examples of the different
types of classical extended radio galaxies, seen at milliJansky and
Jansky levels.  Only one extended AGN has a classic FRII structure,
source 212.  Source 503 is closer to an FRI than an FRII (the small
component the north is source 507), and Source 304 is intermediate
type with well-resolve hot spots.  Source 549 is too weak to determine
if it is FRI or FRII.  The other three sources are FRI, including
Source 69 that does not have hot spots but a strong symmetric jet.

   A potential extended AGN could be comprised of sources 254 and 255.
They are separated by $7''$, with a bright galaxy between them.  We
classified them as independent radio sources because one of then is
identified with a faint optical counterpart. Source 266 has a strong core
with a possible weak jet emanating to the east.  Sources 386 and 391
could possible be associated with a galaxy between them.  Another
uncertainty concerns the compact source 507 near the northern tip of
the extended AGN, source 503.  Because source 507 is clearly
coincident with a galaxy, it has been cataloged as an independent
object.

The relative numbers of extended AGN found in the SSA13 field
are somewhat sparse as expected from other surveys at somewhat higher
flux densities \citep{jac04}.  At about 1 mJy, the density of FRI
sources is about four times that of FRII sources.  This ratio is
difficult to determine from our sample, but is consistent with 25\% of
them as FRII or primarily FRII.  While the extended radio galaxies
comprise about 30\% of the sources at about 1 mJy, only about 10\% of
the extended sources are larger than about $8''$, according to the
SIRTF FLS VLA survey\footnote{http://www.cv.nrao.edu/sirtf\_fls}.  If
this ratio holds for the microJansky population, we would expect that
out of a sample of 510 sources that approximately 15 should be
extended radio galaxies.  This is considerably larger than the five
extended AGN that we detected in the complete sample.

\subsection {Faint Identifications: ROM=F and Unidentified Sources: ROM=U}

   Most of the faint identifications are secure with better than
$90$\% confidence.  This is because the registration accuracy of the
radio/optical images is accurate to $0.2''$, so that proximity of the
radio source to even a 25.5-mag galaxy (where the mean separation of
objects is $5''$) is sufficient for a confident identification.
Identifications that are less secure than 90\% are: 141, 159, 171,
177, 234, 267, 275, 434, 449, 589, 685, 715, 721, 748 and 754.  Thus,
considering all these less secure identifications as unidentified would
only increase the number to 55 sources or 11\% of the sample.

\subsection {Comparison with the Hubble Deep Field North}

    The only other sensitive observations currently available with
sufficient resolution to compare detailed radio and optical
morphologies are the combined MERLIN/VLA observations of the Hubble
Deep Field North \citep{mux05} with $0.2''$ resolution and 92 sources
in the complete sample (see \S 7.1).  The variety and relative numbers
of the different radio/optical morphologies are similar in both
fields, with most HDF identifications associated with relatively
isolated galaxies.  Although many sources in the HDF sample have radio
emission coincident with the galactic nucleus, there is often
additional radio emission with an angular extension that is correlated
with the galaxy orientation.  Slight displacements of the radio
centroid from the nucleus are clearly observed in both the HDF and
SSA13 fields, and in both surveys, approximately 10\% of the sources
have radio emission which is well-displaced from the galaxy nucleus.

    These surveys show that with arcsecond or sub-arcsecond resolution
both the optical and radio emission brightness distributions are
complex in a significant part of the microJansky source population.  Not
only can solid identifications be made, but a detailed comparison and
modeling of the radio and optical brightness distributions can lead to a
deeper understanding of the role of both AGN and starburst activity in the
evolution of galaxies.

\section {Radio Properties}

\subsection {Angular Size Distribution}

The size of each radio source is given in columns 7, 8 and 9 of Table
2.  An unbiased distribution of the true angular size is difficult to
determine for at least two reasons: the inability to measure the
angular size is a strong function of SNR, and the definition of what
constitutes a single radio source is ambiguous.  With these caveats,
Fig.~9a shows the number distribution of the largest angular size
(LAS: the major axis size or separation in the cases of
multi-component sources) of the 289 sources in the complete sample
which have SNR$>8$, equivalent to sources with $S_{1.4}>40~\mu$Jy/beam
at the field center.  With this high SNR, source diameters larger than
$1.2''$ can be reliably determined, and we find that about 64\% (186
of 289) of the sources are smaller than this angular size.

      Five of the 13 sources larger than $4''$ in the SSA13 field are
extended active galaxies (212, 304, 446, 503 and 621).  Two sources
(72 and 758) are associated with a bright flattened galaxy with a
large radio disk component; three sources (330, 385 and 629) are
associated with several galaxies in a tight group of galaxies; two
sources (188 and 517) have diffuse radio emission associated with one
galaxy.  Source 711 is associated with a binary galaxy and it could be
interpreted as two independent sources.  About 20 other radio sources
have hints of extended structure larger than $4''$, but are too faint
to be reliably measured.

    Fig.~9b shows the angular size distribution obtained by the
combined MERLIN/VLA observations of the Hubble field with a resolution
of $0.2''$ \citep{mux05}.  Although there are only 92 sources in this
sample, virtually all of them are resolved and have a measured largest
angular size.  The proportion of sources with LAS small than $1.2''$
is similar for the SSA13 and HDF fields, and most of the remaining
sources are smaller than $4''$.  The median angular size is about
$1.2''$, but there are two parts to the angular distribution: a
Gaussian shaped distribution centered near $0.7''$ (resolved only by
the MERLIN observations), and a tail which decreases roughly as the
inverse angular size.

There have been several investigations on the number of `missing'
large-diameter, microJansky sources in VLA catalogs
\citep{gar00,mux05}.  The Westerbork surveys and the SIRTF FLS surveys
with $\approx 10''$ resolution find that about 15\% to 20\% of the
sources at the $300~\mu$Jy level are larger than about $4''$,
whereas only 4\% of the sources from our SSA13 catalog are at least
this large.  But, when observations with higher angular resolution are
available, a significant fraction of this excess of apparently large
sources found in low resolution observations is explained by the
blending of weak, close sources which thus appear as a single resolved
source.  This type of ambiguity is also seen for a few of the
cataloged sources in the SSA13 field.  Nevertheless, it is possible
that some relatively large and diffuse sources are missing (resolved
out) in our higher resolution radio images.  We thus conclude that
about $8\pm 4$\% of the microJansky radio sources are larger than
$4''$ in angular size.  The upper limit nearly reaches the proportions
seen in lower-resolution surveys, and the lower limit is about that
observed in the SSA13 field.

\subsection {Spectral Index}

    Observation of a small solid angle near the central part of the
SSA13 field have been made at 8.4 GHz \citep{fom02}.  Within a radius
of $273''$, 34 sources were detected above an image level of
$7.5~\mu$Jy/beam (5-$\sigma$).  In order to determine the spectral
index of a complete sample defined at 1.4 GHz, we have examined the
field within $180''$ from the field center where the rms sky noise at
8.4 GHz is $<4.3~\mu$Jy/beam and at 1.4 GHz is $<5.5~\mu$Jy/beam.  In
this region, 47 sources are detected at 1.4 GHz in the complete
sample, all of which are seen on the 8.4 GHz image, above the
2.5-$\sigma$ level.  Although the number of measured spectral indices
is small, the detection levels at 1.4 GHz and 8.4 GHz were
well-matched so that the strong bias in the derived spectral index
distribution that can occur because of a non-detection at one of the
frequencies is not present in this sample.

    The spectral index histogram ($S\propto\nu^\alpha$) for the
complete sample of 47 sources at 1.4 GHz is shown in Fig.~10a.  The
mean spectral index is $-0.84\pm 0.08$.  For comparison, the spectral
index distribution based on the complete sample of 34 sources at 8 GHz
\citep{fom02} is shown in Fig.~10b. The average spectral index is
$-0.68$ with a standard deviation of the mean of 0.08.

    Fig.~11 shows the radio contour/optical images of the six sources
with the flattest ($\alpha>-0.5$) and the steepest ($\alpha<-1.1$)
spectral indices.  Some possible differences are: 1) The flat-spectrum
spectrum sources are identified with relatively isolated galaxies (434
is probably unidentified) that are relatively undistorted; whereas the
steep-spectrum sources tend to have more complex optical emission.  2)
The flatter-spectrum sources contain a dominant component with an
angular size less than $0.9''$, whereas the steep sources are both
large and small.  3) The average (r-z) magnitude color difference for
the flat-spectrum sources is 1.8, compared with 0.7 for the steep
sources. With such a small sample, this difference is not significant.

We have divided the 1.4 GHz sample, based on total flux density at 1.4
GHz, into those stronger than $75~\mu$Jy (19 sources) and those
weaker than $75~\mu$Jy (28 sources), and these distributions are
shown in Fig.~12a and b.  The mean spectral index for the stronger
sources is $\alpha=-0.78\pm 0.04$ and for the weaker sources is
$\alpha=-0.87\pm 0.05$.  Although both distributions contain several
steep and flat spectrum objects, the peak of the spectral indices is
slightly flatter for the stronger sources.  The probability that these two
distributions come from the same population is 20\%.

\subsection {The Counts of Radio Sources}

With this survey we are able to determine the density of radio sources
in the sky from $26~\mu$Jy to about $2000~\mu$Jy. Table 4 shows the
observed source counts as a function of sky flux density for the 560
sources in the complete sample.  Columns 1, 2 and 3 give the source
total flux density binning information, and column 4 gives the number
of sources found within each bin.  Column 5 gives the effective weight
of the data in this bin.  A weight of 15.0 for the lowest flux density
bin, for example, means that on average a source within this flux
density range would have a peak image flux density greater than the
completeness threshold of $25.8~\mu$Jy/beam in only 1/15.0 of the
$15'$ radius region.  The determination of the weighting factor is
described in the following paragraph.  Columns 6 then gives the
derived integral count and column 7 gives the differential count for
each flux density bin.  In Column 8 the differential count is
normalized to a Euclidean count of $1\times S^{-2.5}dS~$sr$^{-1}$
Jy$^{-1}$.  The errors in the count are derived from the number of
sources in each bin.

     The weight in column 5 corrects for the smaller area in which
weaker sources can be detected above the image detection
(completeness) level.  The weight depends strongly on the radial
dependence of observational sensitivity over the field of view, shown
in Fig.~2, but also takes account of the angular size distribution of the
microJansky sources (see Fig.~9) and the statistical noise of the image.
Simulations of a sky filled with point sources, then convolved with
the above effects, were used to determine the weights as a function of
flux density for each source.

   The derived normalized counts for the SSA13 field, from Column 8 in
Table 4, is plotted in Fig.~13.  We have also included the counts from
other observations of comparable sensitivity at 1.4 GHz.  References
to the other data are: PDFS and PDF \citep{hop98}, MC85 \citep{mit85},
OW84 \citep{oor85}, ELIAS \citep{cil00} and HDF \citep{ric00}).  The
SSA13 field has a somewhat higher density of sources than other fields
at these flux densities, particularly the HDF field which has only
60\% of the source density seen in SSA13.

The field to field apparent differences in the counts at and below the
one milliJansky level have been predicted \citep{ben95} and previously
noted \citep{hop98}.  With a typical field size of one degree, the
inhomogeneity between the counts of various deep fields corresponds to
a fluctuation volume scale size of order 100 Mpc$^3$.  Some of the
variations in the counts can be caused by various instrumental
defects; however, there are a sufficient number of deep survey using
the same instrument and observational set-up that show significantly
different counts; such as the SSA13 and HDF fields.  The observed
differences in these counts in this case reflects real cosmic
variance.

   The best fit power-law to the SSA13 alone (neglecting the lowest
flux density bin between $25.8~\mu$Jy and $37.5~\mu$Jy which
may be incomplete), is shown by the dashed line in Fig.~13, and can be
expressed in several ways.  The differential count to the best fit
power law is $n(S)dS = (9.2\pm 0.8)S^{(-2.43\pm 0.13)}$
sr$^{-1}$Jy$^{-1}$.  The integral form of this best fit is
\begin{equation}
   N(>S) = (0.40\pm 0.04)\Big(\frac{S}{75~\mu\hbox{Jy}}\Big)^{(-1.43\pm 0.13)}
\end{equation}
where N is the number of sources (arcmin)$^{-2}$ with a flux density
$>S (\mu$Jy).  This is consistent with the Euclidean value of $-1.50$.

There is some indication that the density of sources is beginning to
decrease below about $48~\mu$Jy, since the count in the
$37.5~\mu$Jy to $48.0~\mu$Jy bin is low.  However, the
decreased density observed near the lower flux density limit of the
count can be caused by a slight error in the weight of the sources, or
an underestimated rms noise near the center of the image where the
blending of faint sources and image artifacts may be occurring (a
minor fall-off at low flux is also seen in the HDF counts).
Extrapolation of the best fit slope of $-2.43$ to one microJansky
gives a source density of 148 sources (arcmin)$^{-2}$, which
corresponds to a mean separation between sources of $5''$.  However,
the true separation could be a as much as a factor of two larger if
the count of sources does in fact decrease below about
$48~\mu$Jy.

    Fig.~14 shows the comparison of the source counts at 1.4 GHz and
at 8.4 GHz \citep{fom02} for the SSA13 field.  The slope of the 8.4
GHz counts, $\gamma=2.11\pm 0.18$, is considerably different from the
1.4 GHz value, $\gamma=2.43\pm 0.13$. although the
difference is compatible with the change of the spectral index
distribution at $S_{1.4}\sim 75~\mu$Jy.  To show that this change of
the slope in the count is consistent with the spectral index versus
flux density relationship shown in Fig.~12, we assumed the best
power-law fit at 1.4 GHz (the top line in Fig.~14) and convolved it
with this spectral index distribution.  The dashed line in Fig.~14 is
the predicted count at 8.4 GHz, and it is in good agreement with that
observed.  Thus, the source count change between 1.4 GHz and 8.4 GHz
supports the steepening of the spectral index at 1.4 GHz for sources
fainter than about $70~\mu$Jy.

\section{Summary}

We present a catalog of over 800 radio sources in the SSA13 field,
derived from VLA observations at 1.4 GHz.  About 90\% of the sources
in the complete sample are identified with an optical counterpart on
Subaru r-band or z-band images.  Even with conservative assumptions at
least 88\% of the sources are reliably identified.  Table 2 lists the
radio and optical parameters for each source.  In Fig.~6, radio
contours are overlayed on optical false-color displays for all sources
in order to display the morphological properties of the radio/optical
emission.  This catalog, especially the 560 sources in the complete
sample, can be used as a basis for further studies of the nature of
faint radio sources.  For example, the evolution of ultra-luminous
infrared galaxies has been investigated using data from this survey
\citep{cow04}

      With the relatively high radio and optical resolution and
sensitivity of this work, the complexity of the radio/optical
morphologies has become evident.  We analyzed the morphological
properties into a small number of classifications that depend on the
radio emission size, the complexity of the optical region, and the
relative location of the radio and optical emission.  More than half
of the radio sources are identified with isolated galaxies, and about
30\% with binary or more complex optical systems.  Only seven sources
are identified with FRI or FRII active galaxies, and eight stellar
objects (half of which are blue and are probably quasars) have been
detected.  The centroid of the radio emission is significantly
displaced from the galaxy nucleus for about 8\% of the sources.  The
radio orientation, when resolved, is often similar to that of the
galaxy or binary system, and suggests that the radio and optical
emission are associated with extended, starforming regions.

      The radio spectral index between 1.4 and 8.4 GHz steepens for
sources fainter than about $75~\mu$Jy, and this trend is confirmed by
the different slopes found in the source counts at these two
frequencies.  The count of microJansky radio sources at 1.4 GHz is
close to the Euclidean value down to $50~\mu$Jy.  Although the
observed count appears to fall off below about $40-50~\mu$Jy, this may
be the result of incompleteness at the faint end of the
catalog.
     
    The proportion of sources which are dominated by AGN or by
starforming mechanisms can not be well-determined without more
detailed observations of the sources at other wavebands, but both
mechanisms are clearly present.  Extended radio galaxies ($>4''$),
quasars and galactic stars comprise only a few percent of the sample.

    The observed radio spectra reflect the decreasing numbers of AGN
at flux densities below $100~\mu$Jy.  Although the proportion of
sources with a spectral index $>-0.5$ is about 10\% over all flux
density ranges, there is a clear increase in the number of
steep-spectrum sources ($\alpha<-0.9$) at flux densities below
$75~\mu$Jy.  The emission from these weaker sources is more likely
caused by remnant plasma from starburst and supernova phenomena, than
by AGN-induced jets and lobes.

    The radio/optical morphology is also a useful discriminant between
emission from starbursts and/or AGN activity.  From Table 3,
approximately 40\% of the radio sources contain radio emission that
is located in the galactic nuclear region and is less than $<1.5''$
in extent.  Although these sources could be regions of dense star
formation near the nuclear regions as in Arp220 \citep{ana00}, these
statistics at least provide an upper limit of 40\% to the fraction of
sources which could be AGN cores. The angular size of the radio emission is
not a definite discriminator between AGN and starburst phenomena
unless resolutions better than $0.05''$ can be reached.  Recent
VLBA+GBT observations of the Bootes field \citep{gar01,gar05} show that
approximately 8\% of the sub-mJy sources have appreciable flux density
in a components with a brightness temperature greater than $10^5$ K;
whereas 29\% of the sources stronger than 1 mJy have such a bright,
non-thermal component.

    With the high sensitivities and resolutions now available fom
radio, optical and X-ray observations, it is already clear that most
galaxies formed at early cosmological epochs show complexity in
emission mechanisms and morphological structure.  These are only now
being probed with the most recent observations, which may lead to a
deeper understanding of AGN formation and the starburst
phenomenon, and the interaction between them.

\section {Acknowledgments}

    The National Radio Astronomy Observatory is a facility of the
National Science Foundation operated under cooperative agreement by
Associated Universities, Inc.  EAR was supported by a Hubble
Fellowship and RBP was supported by the NSF grant AST-0071192 to
Haverford College.  We thank the referee for the careful reading and
significant improvements in the text.

\clearpage

\begin{figure*}
\includegraphics{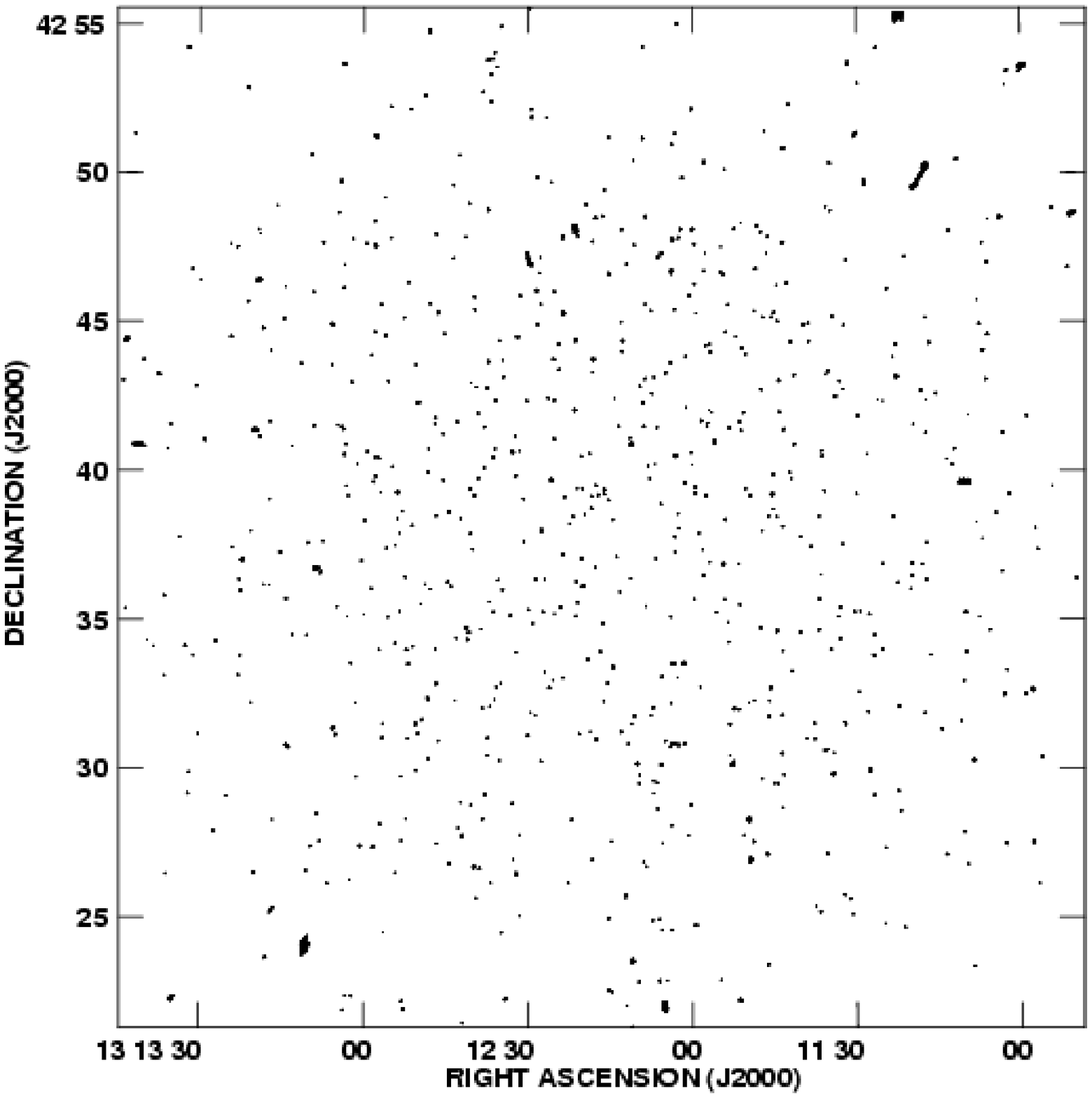}
\vspace{19cm}
\caption []{{\bf The SSA13 1.4-GHz Radio Image.}  The inner $34'\times
34'$ area of the cleaned radio image of SSA13.  The image is {\it not}
corrected for instrumental distortions and the decreased sky
sensitivity as a function of distance from the field center (see
Fig.~2 for the dependence of the sensitivity versus radial distance
from the field center).  All of the dark spots are sources above the
detection level.  A few extended sources are visible.}

\end{figure*}
\clearpage

\begin{figure*}
\includegraphics[scale=0.95]{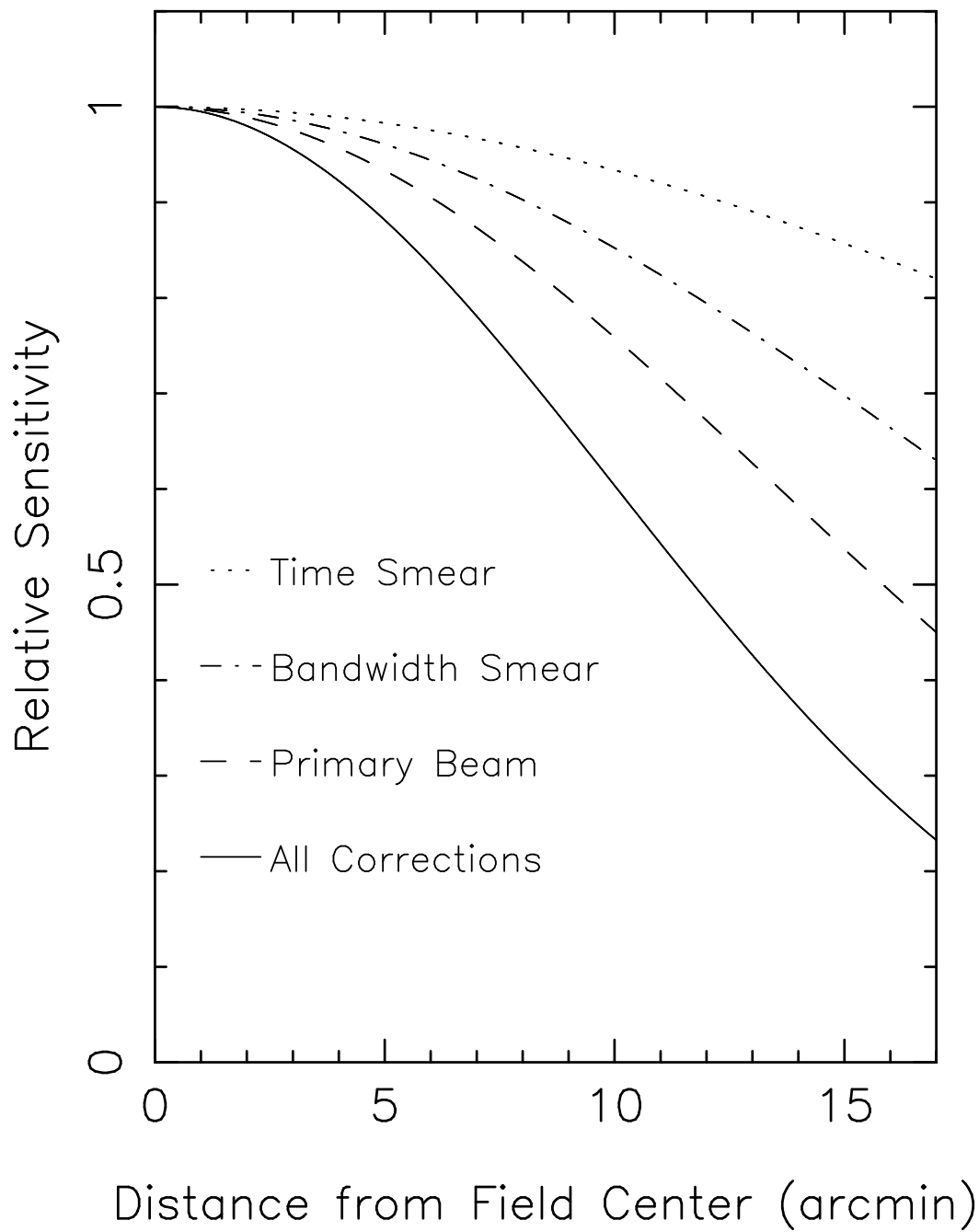}
\vspace{0cm}
\caption {{\bf The Relative Sensitivity Across the Radio Image.} The
ordinate shows the ratio (sky sensitivity / image sensitivity) associated
with the peak intensity of a radio source as a function of radial
distance from the field center.  The fractional sensitivity loss due
to the three factors is discussed in the text, and the net
sensitivity is shown by the solid line.}

\end{figure*}
\clearpage

\begin{figure*}
\includegraphics{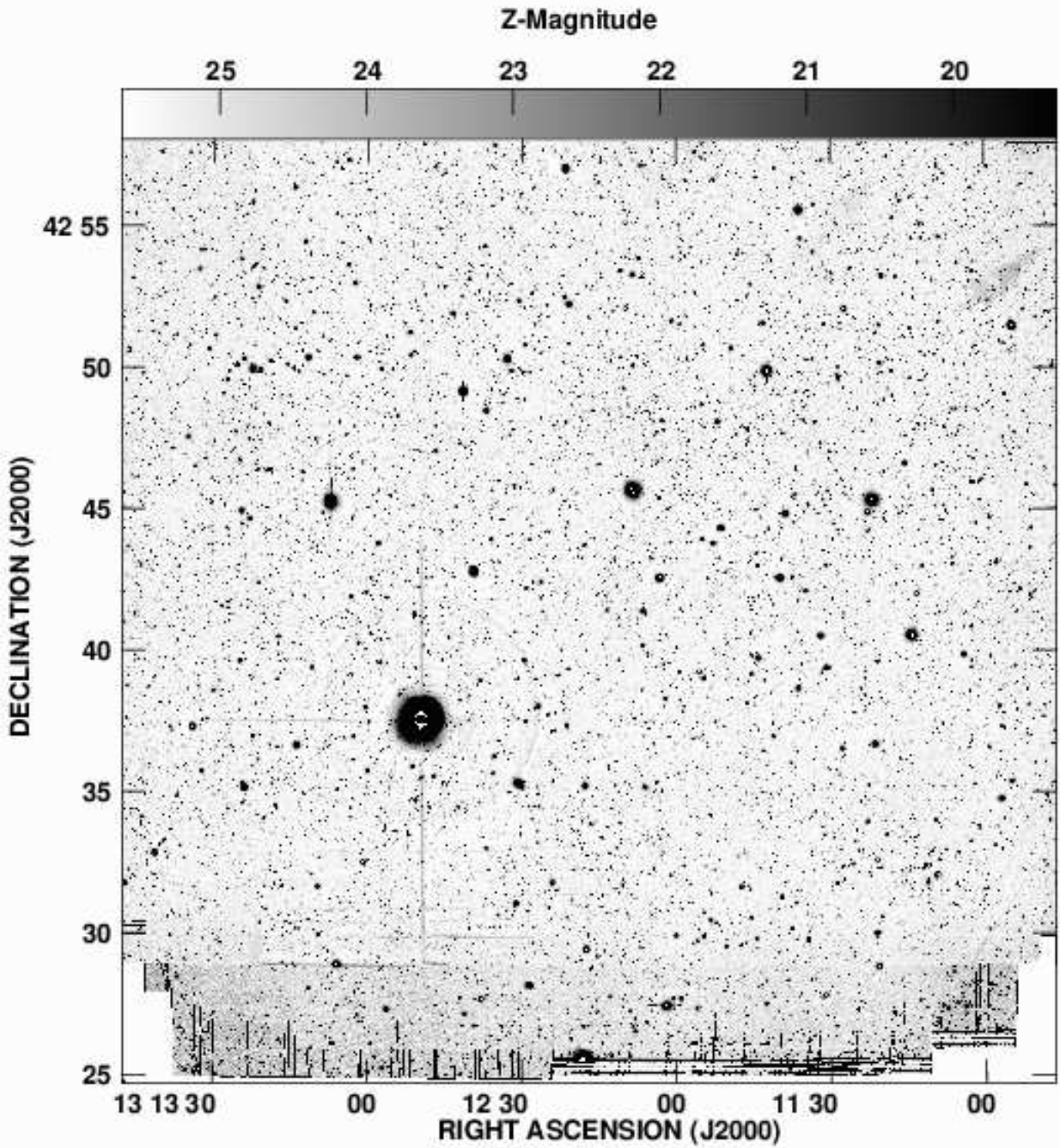}
\vspace{16cm}
\caption {{\bf The SSA13 z-Band Image:} A gray-scale representation of
the entire z-band image.  The contrast,
expressed in magnitudes, is shown by the wedge above the image.}

\end{figure*}
\clearpage

\begin{figure*}
\includegraphics{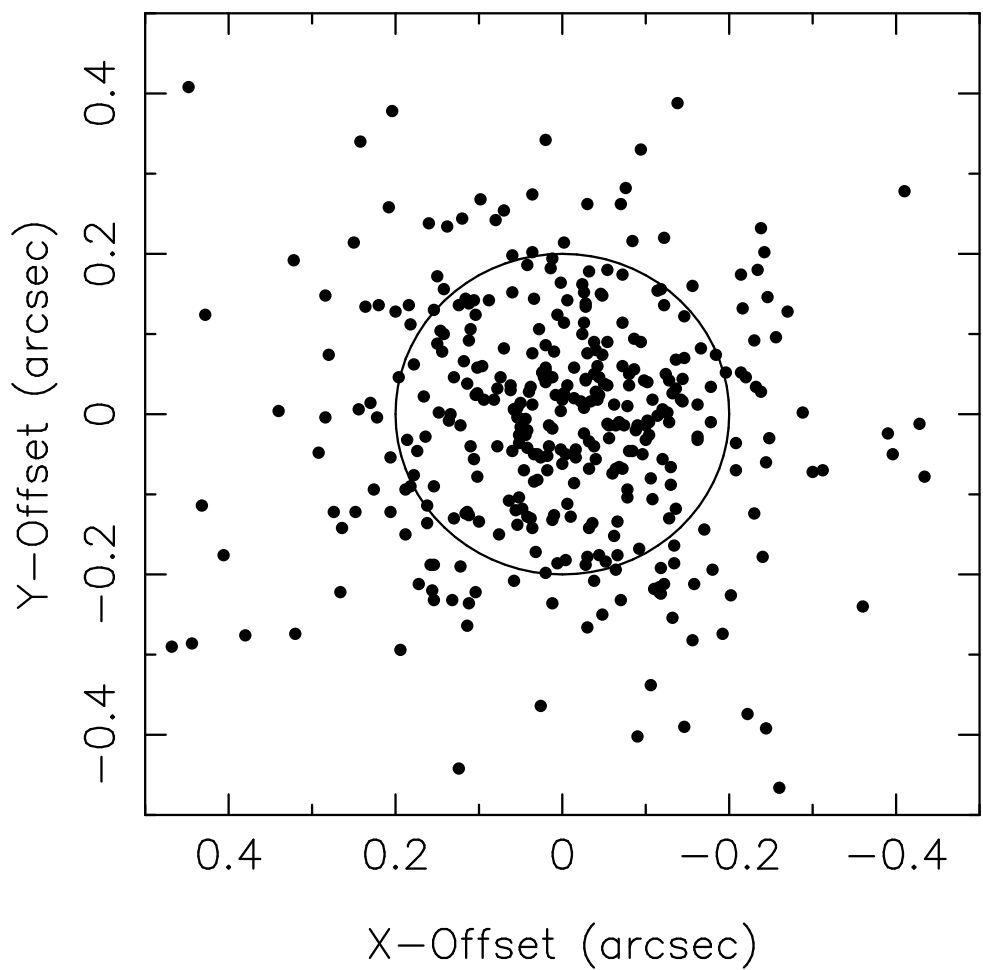}
\vspace{14cm}
\caption {{\bf Radio-Optical Registration:} The difference
between the radio and optical positions for 95 high-quality
identifications.  The circle indicates the one-sigma error of
$0.2''$.}

\end{figure*}
\clearpage

\begin{figure*}
\includegraphics{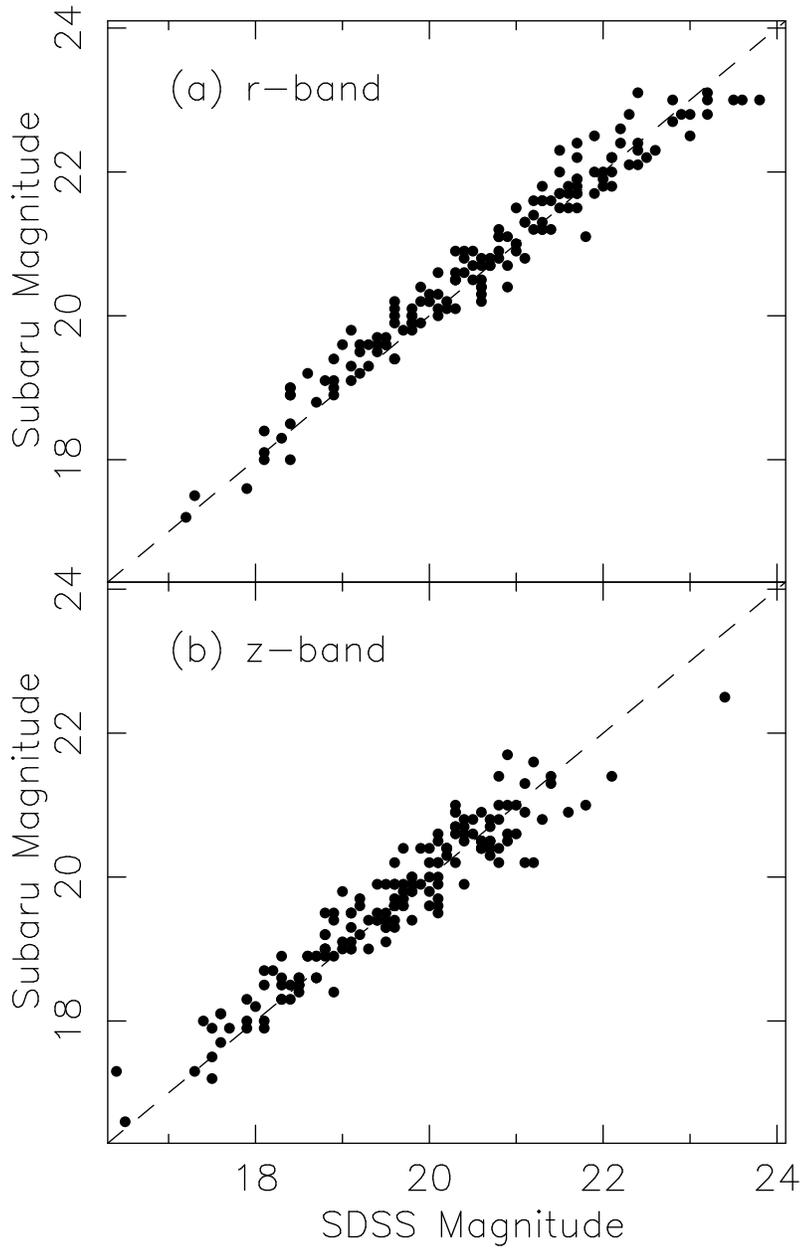} 
\vspace{15cm} 

\caption {{\bf Subaru vs SDSS Magnitude Comparisons:} (a) The
comparison of Subaru and SDSS r-magnitudes for 110 galaxies.  (b) The
comparison of Subaru and SDSS z-magnitudes for 113 galaxies.  The
comparison is made after zero-point correction of the Subaru
r-magnitude by 0.1 mag and z-magnitude by 0.4 mag.}

\end{figure*}

\clearpage

\begin{figure*}
\includegraphics{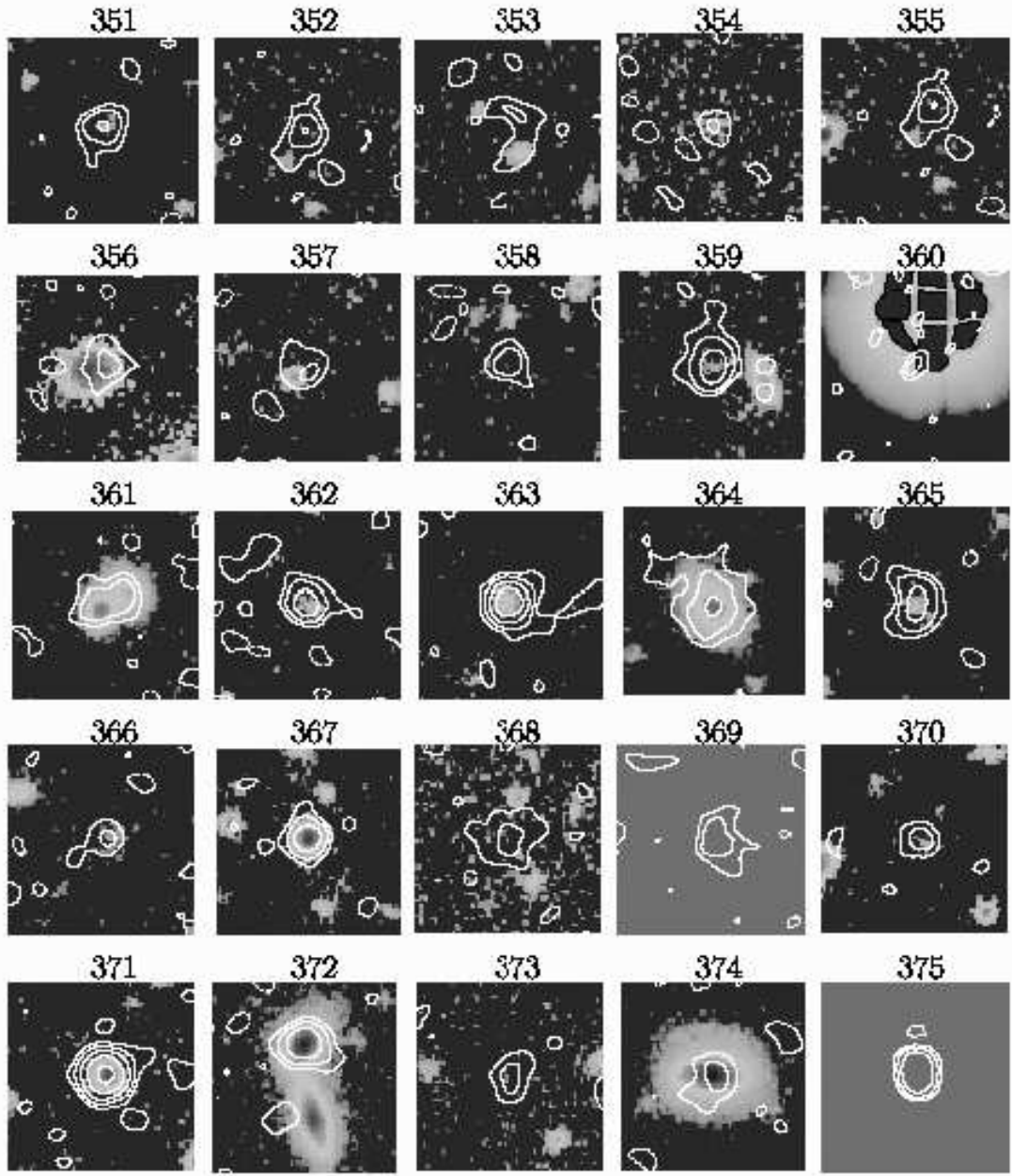}
\vspace{16cm} 
\caption{{\bf The Radio/Optical Images:} The catalog number is shown
above each diagram.  The contours show the radio emission taken
directly from the
image of Fig.~1 (that is, uncorrected for the decrease of sky
sensitivity with linear distance from the field center).  The lowest
radio contour value is $10.0~\mu$Jy/beam and levels are shown
at -1,1,2,4,16
times the lowest value (except for the brightest sources where the
lowest contour level has been increased).  The optical emission
(usually z-band) is shown by the false-color image, adjusted to
display the optical morphology.  Bright sources (360) can be saturated.
A uniform pink background is used
when no Subaru optical data are available.
{\bf NOTE: All 33 pages of color images can be downloaded from
ftp://ftp.cv.nrao.edu/pub/NRAO-staff/efomalon/ssa13\_diagrams/}}

\end{figure*}
\clearpage

\begin{figure*}
\includegraphics{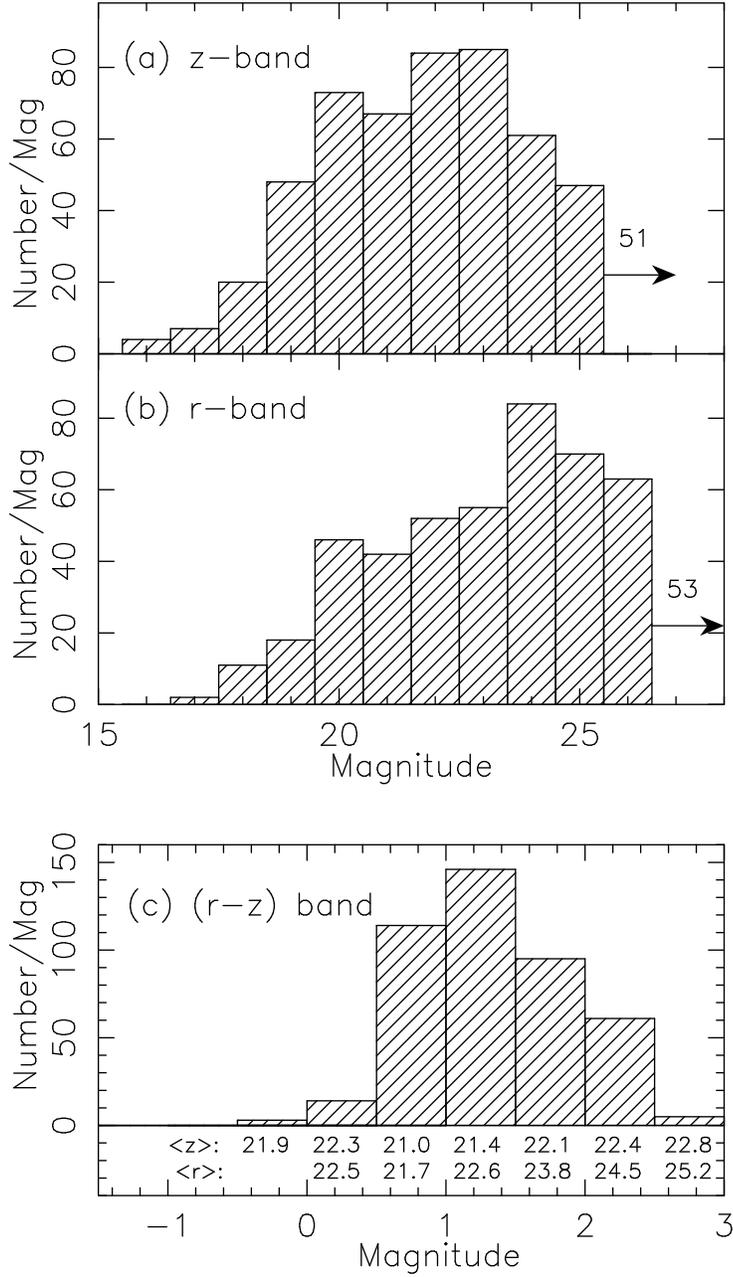}
\vspace{17cm}
\caption{{\bf Magnitude Distributions for Optical Identifications:}
(a) The distribution of the z-band identifications for the the 532
sources in the complete sample with Subaru z-band data.  Fifty-one
sources are not identified.  (b) The distribution of the r-band
identifications for the 480 sources in the complete sample with Subaru
r-band data.  Fifty-three are not identified; (c) the distribution of
the r-z color distribution for 441 sources with both r-band and z-band
measurements.  The average r-band and z-band magnitude is listed under
each color magnitude bin; note the trend in $<r>$.}

\end{figure*}

\clearpage

\begin{figure*}
\includegraphics{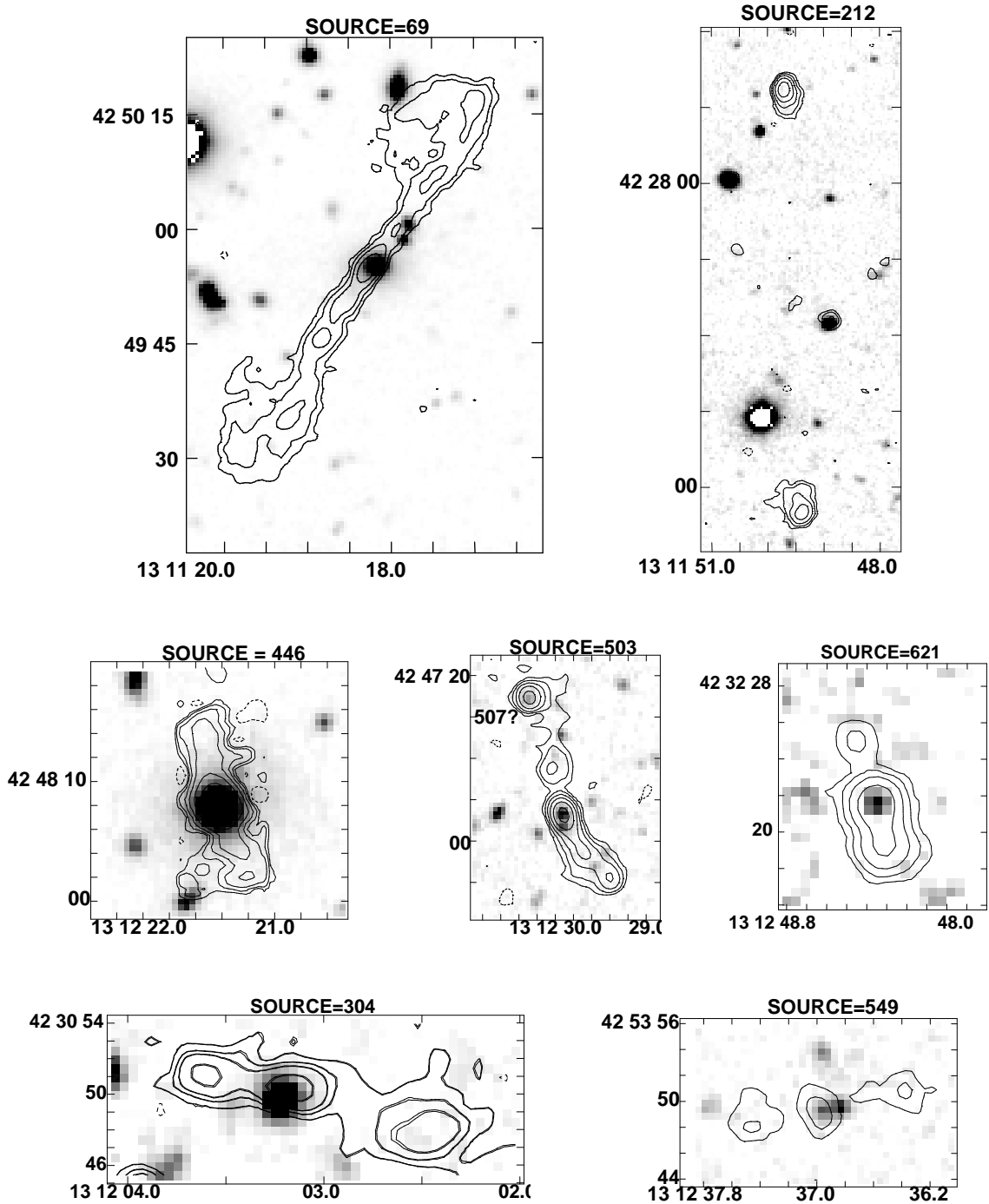} 
\vspace{19cm}

\caption {{\bf Extended Active Galaxies:} The uncorrected radio
contours, from Fig.~1, superimposed on the z-band optical gray-scale,
of the seven AGNs in the SSA13 field are displayed.  The source number
from the catalog is given above each plot.  The minimum contour level
of these images is $27.5~\mu$Jy/beam with levels at -1,1,2,4,8,16,...
times the minimum level.  The gray-scale contrast has been adjusted to
best display the radio/optical alignment.}

\end{figure*}
\clearpage

\begin{figure*}
\includegraphics{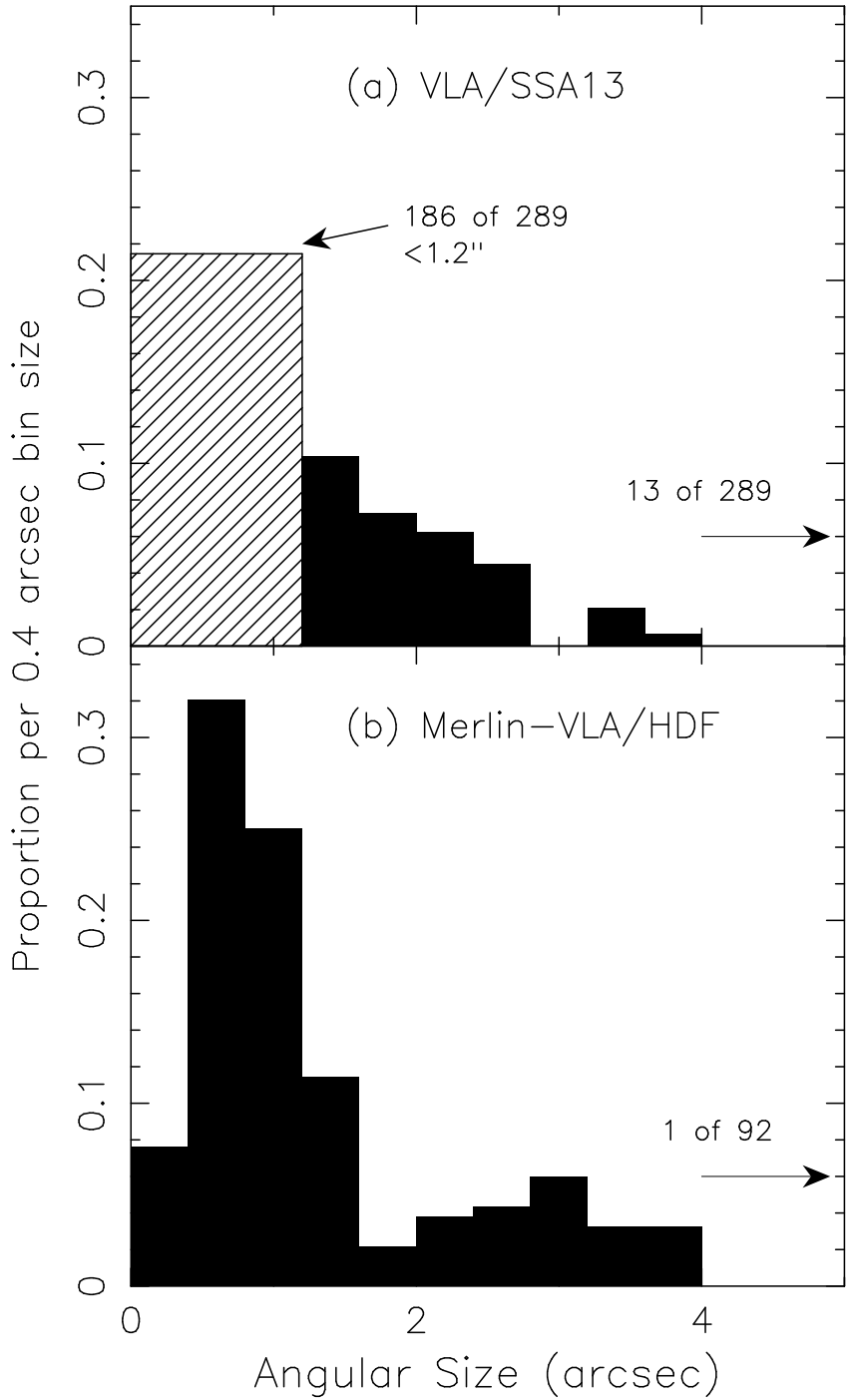}
\vspace{18cm}

\caption {{\bf The Angular Size Distribution at 1.4 GHz:} (a) The
distribution of 289 sources in the complete SSA13 sample having
SNR$>8$.  The unresolved sources, those largest angular size less
than $1.2''$, are shown by the hashed distribution, equally spread
over the angular size range.  Thirteen of the 289 are larger than
$4''$.  (b) The distribution of 92 sources from the combined
MERLIN/VLA observations of the HDF-North field.  All sources are
resolved with the $0.2''$ resolution of this survey \citep{mux05}.
See text about the number of source larger than $4''$}.

\end{figure*}
\clearpage

\begin{figure*}
\includegraphics{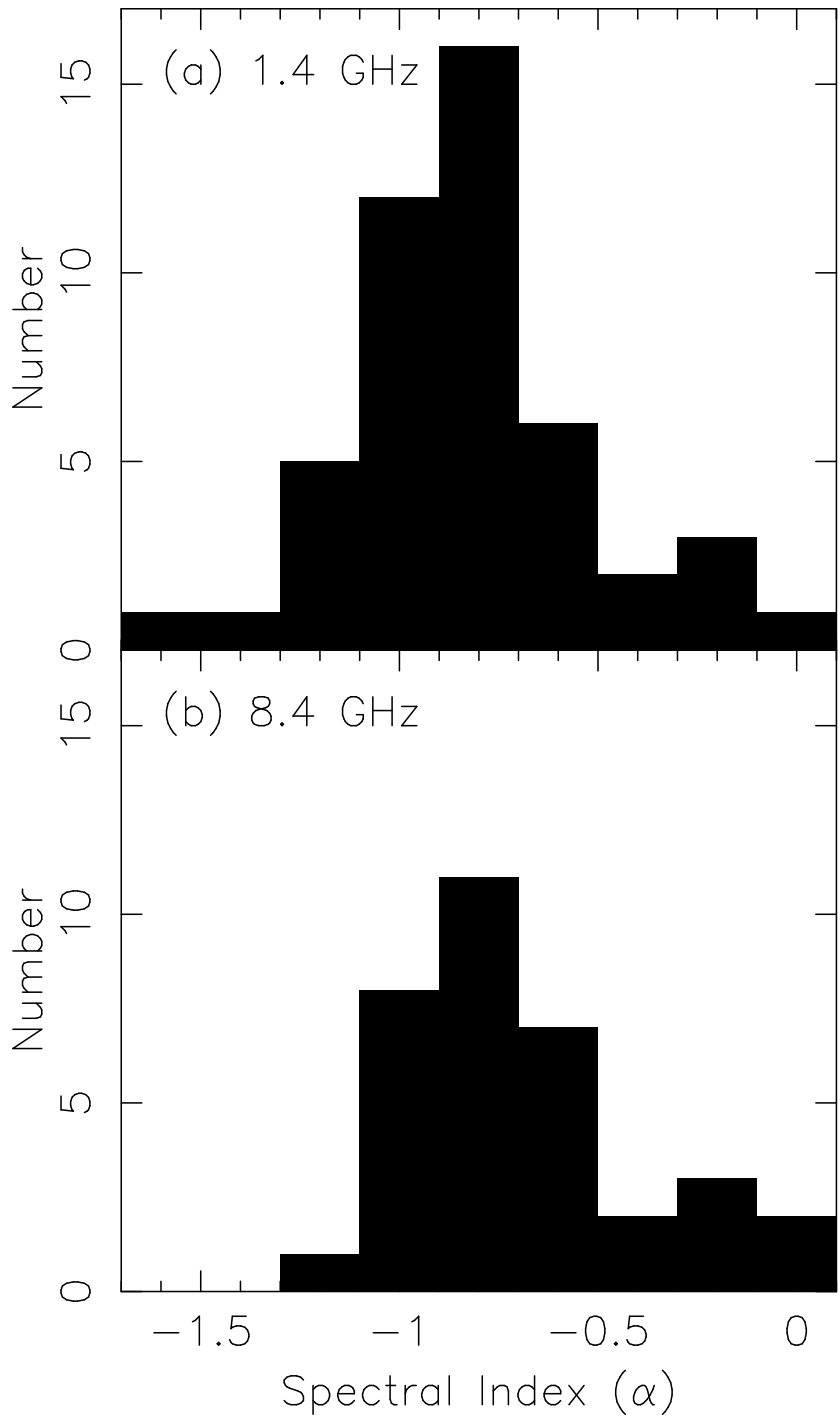}
\vspace{18cm}
\caption {{\bf The Spectral Index Distribution versus Frequency:} (a)
The spectral index distribution of 47 sources in SSA13 from a complete sample
at 1.4 GHz.  (b) The spectral index distribution of 34 sources from a
complete sample at 8.4 GHz.}

\end{figure*}
\clearpage

\begin{figure*}
\includegraphics{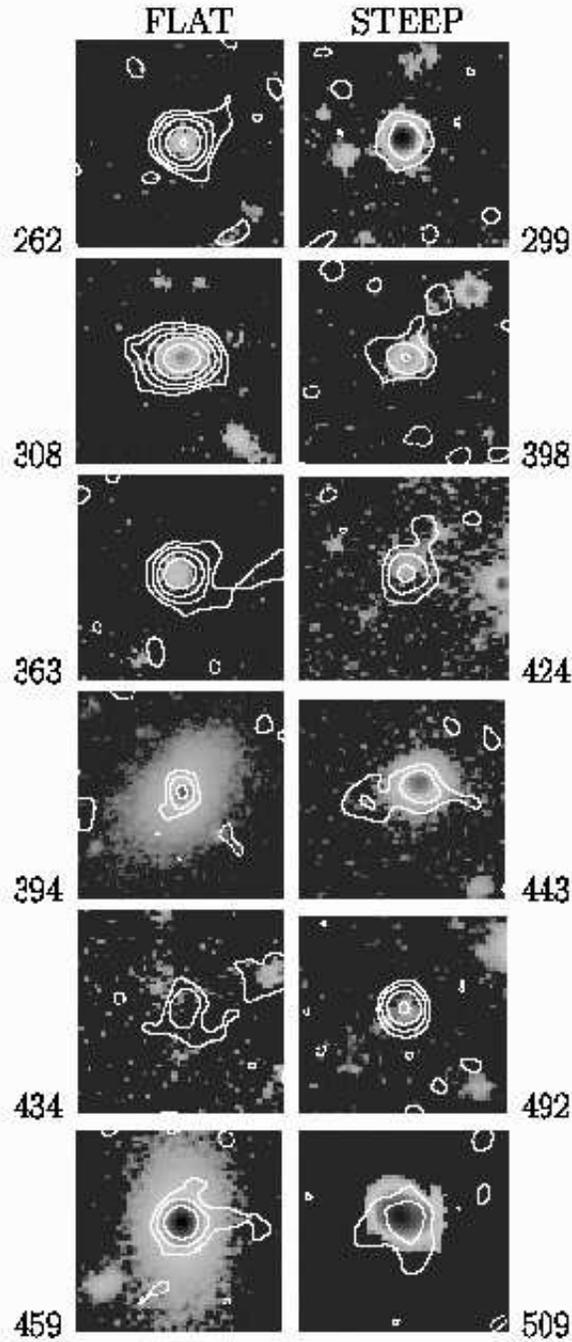}
\vspace{17cm}
\caption {{\bf Sources with Extreme Spectral Indices::}
The images of these sources have been taken from Fig.~6, with contour
levels=-10,10,20,40,160 $\mu$Jy/beam.  The left column shows the
radio/optical structure of the six sources with the flattest spectral
index.  The right column shows the six sources with the
steepest spectral index.  The source catalog number is given at the
lower edge of each diagram.}

\end{figure*}
\clearpage

\begin{figure*}
\includegraphics{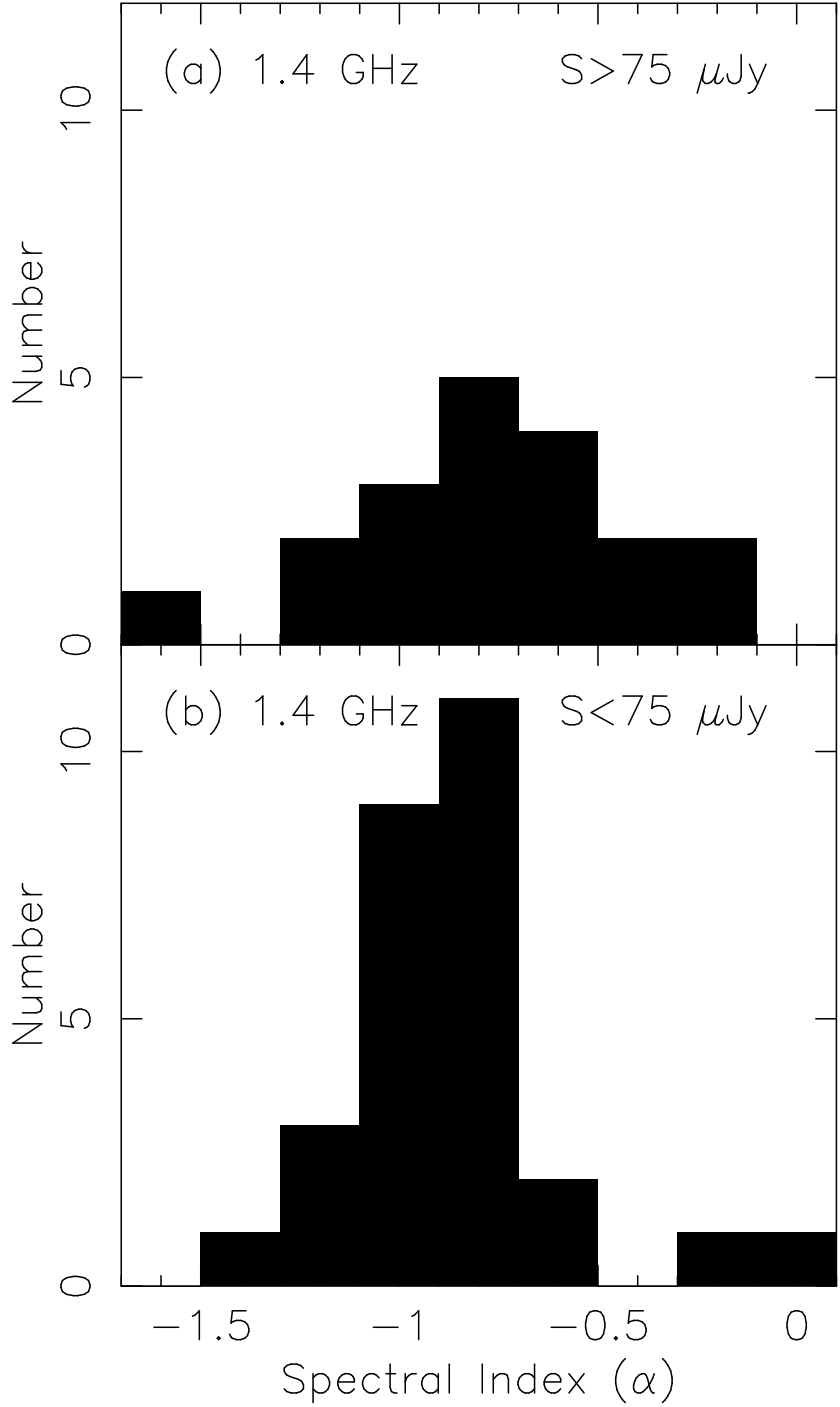}
\vspace{17cm}
\caption {{\bf The Spectral Index Distribution versus Flux Density:}
(a) The spectral index distribution of 19 sources from a complete
sample at 1.4 GHz with flux densities $>75~\mu$Jy.  (b) The
spectral index distribution of 28 sources from a complete sample at
1.4 GHz with flux densities $<75~\mu$Jy.}

\end{figure*}
\clearpage

\begin{figure*}
\includegraphics{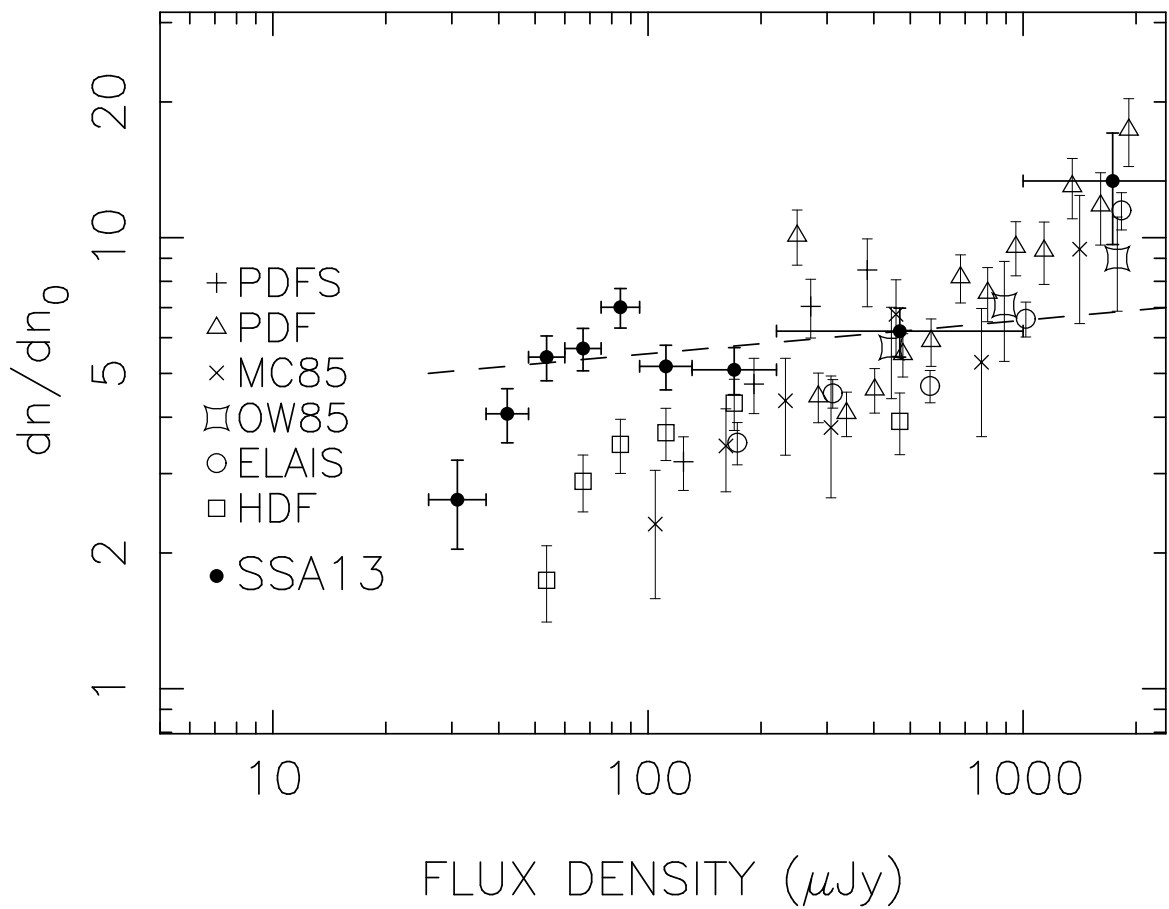}
\vspace{15cm}
\caption {{\bf The Micro-Jansky Counts of Radio Sources at 1.4 GHz:}
The plotted points show the corrected differential count observed
between $25~\mu$Jy and $2000~\mu$Jy at 1.4 GHz from a
variety of deep surveys.  The solid circles are the counts from SSA13.
The dashed line gives the best fit power law to the SSA13 observation,
with the lowest flux density point excluded.  References are given in
the text.}

\end{figure*}
\clearpage

\begin{figure*}
\includegraphics{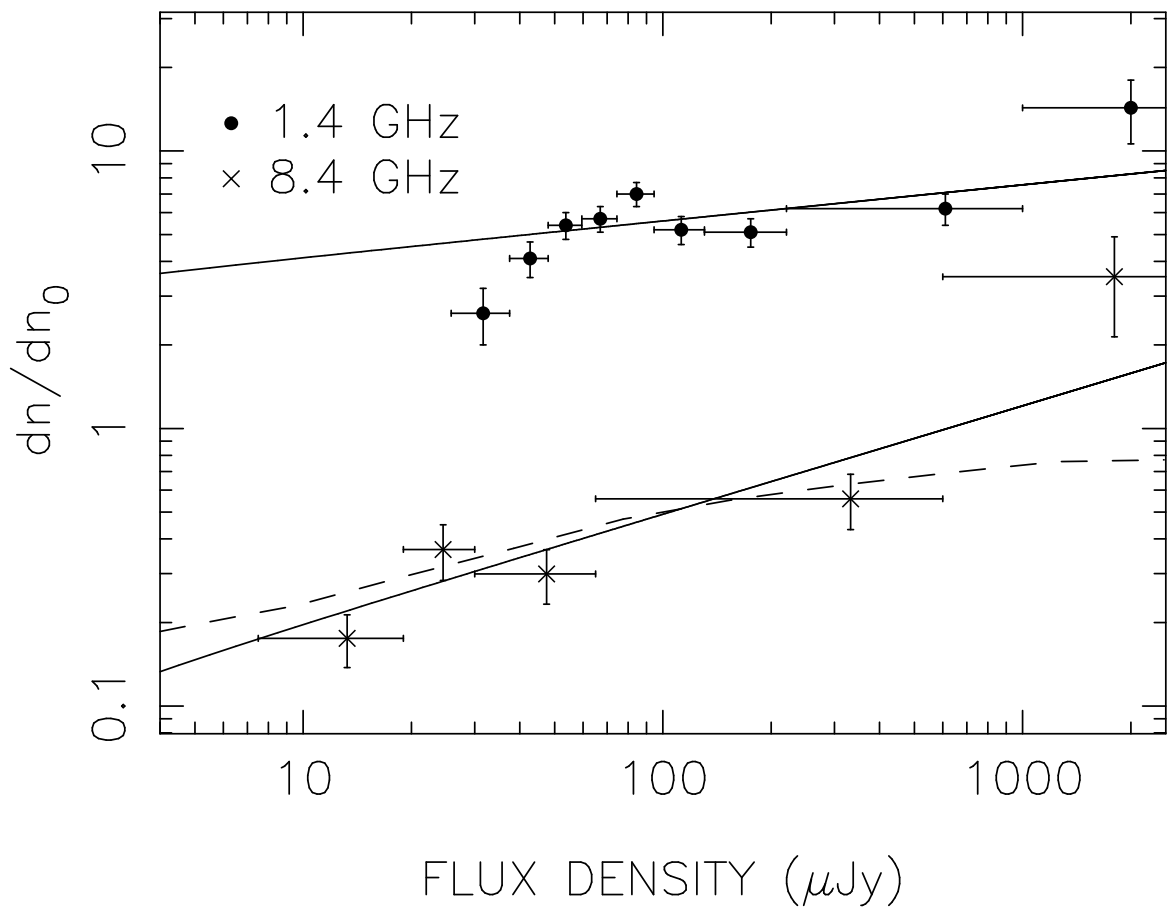}
\vspace{15cm}
\caption {{\bf Conversion of Count Between 1.4 and 8.4 GHz} The 1.4
GHz and 8.4 GHz differential counts from the SSA13 field are shown by
the plotted points, with the solid lines as the best power-law fits.
The dashed curve shows the derived 8.4 GHz source count modeled from
the 1.4 GHz count and the spectral index distribution, shown in
Fig.~12.}

\end{figure*}

\clearpage

\begin{deluxetable}{lll}
\tablecolumns{3}
\tablewidth{0pt}
\tabletypesize{\normalsize}
\tablecaption{SSA13 VLA OBSERVATION LOG}
\tablehead {
 \colhead {Date} &
 \colhead {No. Hours} &
 \colhead {Config.} \\
}
\startdata
1999 July 02   &  11   & A \\
1999 July 04   &  11   & A \\
1999 July 05   &  12   & A \\
1999 July 16   &  12   & A \\
1999 July 17   &   8   & A \\
1999 July 19   &  11   & A \\
1999 July 23   &  13   & A \\
1999 August 30 &  13   & A \\
2000 January 14&  13   & B \\
\enddata
\end{deluxetable}

\clearpage
\begin{deluxetable}{rrrrrrrrrcrr}
\tabletypesize{\scriptsize}
\tabcolsep2.5pt
\tablecaption{Radio/Optical Source Catalog}
\tablehead {
  \colhead{(1)} &
  \colhead{(2)} &
  \colhead{(3)} &
  \colhead{(4)} &
  \colhead{(5)} &
  \colhead{(6)} &
  \colhead{(7)} &
  \colhead{(8)} &
  \colhead{(9)} &
  \colhead{(10)} &
  \colhead{(11)} &
  \colhead{(12)} \\

  \colhead{No} &
  \colhead{Source} &
  \colhead{SNR} &
  \colhead{S$_{1.4}$} &
  \colhead{RA} &
  \colhead{DEC} &
  \multicolumn{3}{c}{Radio Size} &
  \colhead{ROM} &
  \colhead{r} &
  \colhead{z} \\

}
\startdata
N001 & J131047.4+424218 &   4.9 &    132(  35) & 13 10 47.481 (0.033) & 42 42 18.03 (0.17) & $ 2.8$ & $<1.2$ &  92 & Ge & $ 22.7$ & $ 21.2$ \\
N002 & J131048.1+424012 &   6.6 &    149(  36) & 13 10 48.182 (0.020) & 42 40 12.86 (0.17) & $<2.5$ & $    $ &     & Gc & $ 25.0$ & $ 22.7$ \\
N003 & J131049.0+423614 &   3.8 &     77(  29) & 13 10 49.081 (0.028) & 42 36 14.56 (0.27) & $<2.8$ & $    $ &     & U  & $>26.0$ & $>23.4$ \\
N004 & J131049.7+423623 &   7.4 &    131(  32) & 13 10 49.797 (0.016) & 42 36 23.44 (0.19) & $<2.4$ & $    $ &     & F  & $>26.4$ & $ 24.9$ \\
N005 & J131051.7+424308 &   4.3 &     71(  24) & 13 10 51.705 (0.018) & 42 43 08.95 (0.26) & $<2.2$ & $    $ &     & Gd & $ 21.6$ & $ 21.0$ \\
N006 & J131054.2+423929 &   6.6 &     96(  24) & 13 10 54.211 (0.015) & 42 39 29.92 (0.18) & $<1.9$ & $    $ &     & Gc & $ 18.4$ & $ 17.9$ \\
N007 & J131055.7+424346 &   6.1 &     86(  22) & 13 10 55.757 (0.017) & 42 43 46.53 (0.15) & $<1.3$ & $    $ &     & F  & $>26.3$ & $ 23.9$ \\
N008 & J131056.1+423022 &  14.5 &    295(  31) & 13 10 56.153 (0.010) & 42 30 22.28 (0.10) & $<1.3$ & $    $ &     & Bc & $     $ & $ 21.6$ \\
\enddata
\end{deluxetable}

\clearpage

\begin{deluxetable}{lrrl}
\tablecolumns{4}
\tabletypesize{\small}
\tablecaption{RADIO/OPTICAL MORPHOLOGICAL CLASSIFICATIONS}
\tablehead {
  \colhead {ROM} &
  \multicolumn{2}{c}{Number} &
  \multicolumn{1}{l}{~~~~~~~~~~~Description} \\

  \colhead {}  &
  \multicolumn{1}{c}{Entire} &
  \multicolumn{1}{c}{Complete} \\
  \colhead {}  &
  \multicolumn{1}{c}{Catalog} &
  \multicolumn{1}{c}{Sample} \\
}
\startdata
 ALL  &   810~~~~ & 560~~~~ \\
 EG   &     7~~~~ &   5~~~~ &  ~~ACTIVE GALAXIES WITH EXTENDED EMISSION \\
 S    &     8~~~~ &   7~~~~ &  ~~STELLAR OBJECTS \\
 G    &   390~~~~ & 268~~~~ &  ~~ISOLATED GALAXIES, of which \\
 ~~Gc &   194 & 134 &  ~~~~~~Unresolved radio source located near nucleus  \\
 ~~Ge &   166 & 118 &  ~~~~~~Resolved radio source located near nucleus \\
 ~~Gd &    30 &  16 &  ~~~~~~Radio source displaced from nucleus. \\
 B    &   185~~~~ & 138~~~~ &  ~~MULTIPLE SYSTEMS, of which \\
 ~~Bc &    72 &  48 &  ~~~~~~Unresolved radio source located near nucleus of one galaxy  \\
 ~~Be &    91 &  72 &  ~~~~~~Resolved radio source located near nucleus of one galaxy \\
 ~~B2 &    22 &  18 &  ~~~~~~Radio emission from both galaxies of a binary system. \\
 C    &    12~~~~ &  12~~~~ &  ~~COMPLEX SYSTEMS \\
 F    &    98~~~~ &  73~~~~ &  ~~FAINT OBJECTS ($\approx 24.5$ to $26.0$ mag) \\
 U    &    83~~~~ &  52~~~~ &  ~~UNIDENTIFIED \\
 N    &    27~~~~ &   9~~~~ &  ~~NO SUBARU DATA \\
~~sg  &    14 &   2 &  ~~~~~~SDSS identifications \\
~~su  &    13 &   7 &  ~~~~~~SDSS unidentified \\
\enddata
\end{deluxetable}

\begin{deluxetable}{rrrcrccr}
\tablecolumns{8}
\tabletypesize{\normalsize}
\tablecaption{SSA13 RADIO SOURCE COUNT}
\tablehead {
  \colhead {S$_{min}$} &
  \colhead {S$_{max}$} &
  \colhead {S$_{avg}$} &
  \colhead {Num}     &
  \colhead {Wt}      &
  \colhead {n(S$>$S$_{min}$)}    &
  \colhead {dn/dS}   &
  \colhead {dn/dn$_0$}  \\
  \colhead{$\mu$Jy} &
  \colhead{$\mu$Jy} &
  \colhead{$\mu$Jy} & & &
  \colhead {$\times 10^5$sr$^{-1}$}  &
  \colhead {$\times 10^{10}$sr$^{-1}$Jy$^{-1}$} \\
 }
\startdata
  25.8  &  37.5 &  34.1 & 20  &15.0  &  54.3 & 49.3 &  2.6$\pm$ 1.6  \\
  37.5  &  48.0 &  42.4 & 53  & 4.20 &  39.0 & 35.4 &  4.1$\pm$ 2.0  \\
  48.0  &  59.5 &  53.5 & 77  & 2.30 &  31.0 & 25.8 &  5.4$\pm$ 2.3  \\
  59.5  &  74.5 &  67.1 & 86  & 1.54 &  23.2 & 15.5 &  5.7$\pm$ 2.4  \\
  74.5  &  94.5 &  83.8 & 99  & 1.24 &  21.5 & 10.7 &  7.0$\pm$ 2.6  \\
  94.5  & 130.5 & 109.6 & 77  & 1.06 &  14.3 &  4.0 &  5.2$\pm$ 2.3  \\ 
 130.5  & 220.5 & 164.3 & 70  & 1.00 &  12.3 &  1.4 &  5.1$\pm$ 2.3  \\
 220.5  & 999.5 & 419.0 & 65  & 1.00 &  11.4 & 0.15 &  6.2$\pm$ 2.5  \\
 $>999.5$ &       &6400.0 & 13& 1.00 &   0.23 & 0.011& 14.3$\pm$ 3.5  \\

\enddata
\end{deluxetable}

\end{document}